\documentclass[pra,aps,superscriptaddress,superbib,citeautoscript,reprint]{revtex4-1}

\usepackage{amsmath,bm,mathrsfs}
\usepackage{graphicx}

\newcommand{\br}{\mathbf{r}}
\newcommand{\dr}{d{\mathbf{r}}}

\newcommand{\bxi}{\bm{\xi}}
\newcommand{\bzeta}{\bm{\zeta}}
\newcommand{\rmvac}{{\rm vac}}
\newcommand{\ie}{{\it i.e.}}
\newcommand{\myket}[1]{\left\vert #1\right\rangle}
\newcommand{\mybra}[1]{\left\langle #1\right\vert}
\newcommand{\braket}[1]{\left\langle #1 \right\rangle}
\newcommand{\ketbra}[1]{{\left\vert #1 \right\rangle\left\langle #1 \right\vert}}

\newcommand{\cdop}{\hat{c}^{\dagger}}
\newcommand{\be}{\begin{equation}}
\newcommand{\ee}{\end{equation}}

\newcommand{\xinu}{\xi_{\nu}}
\newcommand{\xilambda}{\xi_{\lambda}}

\newcommand{\dxim}[1]{\dfrac{\partial #1}{\partial \xi_{-}}}
\newcommand{\dxinum}[2]{\dfrac{\partial #1}{\partial \xi_{#2}}}

\newcommand{\dxilambda}[1]{\dfrac{\partial #1}{\partial \xi_\lambda}}
\newcommand{\Gammaop}{\hat{\Gamma}}

\DeclareMathOperator{\Tr}{Tr}
\DeclareMathOperator{\sgn}{sgn}

\usepackage[colorlinks=true,urlcolor=blue,citecolor=blue,linkcolor=blue]{hyperref}

\newcommand{\LCQ}{Laboratoire de Chimie Quantique, Institut de Chimie, CNRS/Universit\'{e} de Strasbourg, 4 rue Blaise Pascal, 67000 Strasbourg, France}

\begin{document}

\title{
Extended $N$-centered ensemble density functional theory of double electronic excitations 
}

\author{Filip Cernatic}
\affiliation{\LCQ}
\author{Emmanuel Fromager}
\affiliation{\LCQ}


\begin{abstract}
A recent work [arXiv:2401.04685] has merged $N$-centered ensembles of neutral and charged electronic ground states with ensembles of neutral ground and excited states, thus providing a general and in-principle exact (so-called extended $N$-centered) ensemble density functional theory of neutral and charged electronic excitations. This formalism made it possible to revisit the concept of density-functional derivative discontinuity, in the particular case of single excitations from the highest occupied Kohn--Sham (KS) molecular orbital, without invoking the usual ``asymptotic behavior of the density" argument. In this work, we address a broader class of excitations, with a particular focus on double excitations. An exact implementation of the theory is presented for the two-electron Hubbard dimer model. A thorough comparison of the true physical ground- and excited-state electronic structures with that of the fictitious ensemble density-functional KS system is also presented. Depending on the choice of the density-functional ensemble as well as the asymmetry of the dimer and the correlation strength, an inversion of states can be observed. In some other cases, the strong mixture of KS states within the true physical system makes the assignment ``single excitation" or ``double excitation" irrelevant.
\end{abstract}

\maketitle



\section{Introduction}\label{sec1}

In the mean-field (or noninteracting) description of electronic structures, such as
Hartree--Fock (HF) theory~\cite{slater1930note,fock1930paperonhfmethod} and Kohn--Sham density-functional theory (KS-DFT)~\cite{kohn1965selfconsistent},
a \textit{double excitation} refers to the
promotion
of two electrons from occupied orbitals
in a reference configuration (usually a ground-state Slater determinant)
into two virtual orbitals, resulting in a new configuration, ``doubly-excited" relative to the reference. 
In practice, this simple
picture of distributing
electrons among orbitals in a single configuration is often used as a starting point for describing
\textit{neutral} excitation processes (\textit{i.e.}, processes involving two states with the same number of electrons) in interacting
many-electron systems.
In the latter, doubly-excited configurations alone no longer reflect the full details of the electronic structure of excited states, which are in general described by configuration expansions with single and multiple (double and higher) excitations from the reference~\cite{loos2019reference}. Contributions from the doubles are absolutely essential in many applications, such as the study of 
excited states in conjugated molecules~\cite{lappe2000on,serranoandres1993towards,hsu2001excitation,starcke2006how},
singlet fission~\cite{smith2010singlet,smith2013recent}, and autoionizing resonances~\cite{elliott2011perspectives},
to cite a few examples.

One of the standard and computationally
affordable methods for computing
neutral excitations and excited-state properties 
in molecules and extended systems
is the linear-response time-dependent DFT (TD-DFT)~\cite{runge1984density,casida1995timedependent,casida2012progress,Lacombe2023_Non-adiabatic}.
In linear response TD-DFT,
single excitations are
explicitly encoded in the KS density-density response
function, from which
any true interacting excitation energy (not only single excitation ones) can in principle be retrieved {\it via} the frequency-dependent
Hartree-exchange-correlation (Hxc)
kernel, which relates to the functional derivative of the time-dependent density-functional Hxc potential.
However, the development of
practically applicable and accurate Hxc kernels is 
far from trivial~\cite{maitra2004double,cave2004dressed,Huix-Rotllant2011_Assessment}, and, in the most commonly used 
adiabatic approximation, the frequency-independent ground-state Hxc kernel is employed.
As a result, double and higher excitations are completely absent
from the computed spectra (the reader is referred to Refs.~\cite{elliott2011perspectives,maitra2022double,casida2012progress,Lacombe2023_Non-adiabatic} for more comprehensive discussions on this matter).

Alternatively, a time-independent
and variational approach to excited states that has recently gained an increasing interest is the theory of many-electron ensembles~\cite{Fan1949_On,theophilou1979energy,Hendekovic1982_equi-ensembles,gross1988rayleigh,deur2017exact,yang2017direct,gould2017hartree,gould2018charge,deur2018exploring,           PRL19_Gould_DD_correlation,Fromager_2020,PRL20_Gould_Hartree_def_from_ACDF_th,Gould2020_Approximately,                       loos2020weight,Gould2021_Ensemble_ugly,gould2023local,Gould2023_Electronic,gould2022single,gould2021double,        Cernatic2022,Schilling2021_Ensemble,Liebert2022_Foundation,Benavides-Riveros2022_Excitations,Liebert_2023_An_exact_bosons,Liebert2023_Deriving,ding2024ground}. Ensemble DFT, which extends regular ground-state DFT to
ensembles of ground and (neutral) excited states, was originally introduced by Theophilou~\cite{theophilou1979energy,theophilou_book} for equi-ensembles and then further generalized by Gross, Oliveira and Kohn~\cite{gross1988rayleigh,gross1988density,oliveira1988density}, hence the name TGOK-DFT~\cite{cernatic2024neutral}. Unlike linear response TD-DFT, TGOK-DFT can describe explicitly any
(single or multiple) excitation process, in principle exactly, with essentially
the same computational cost
as a regular ground-state DFT calculation. A single calculation is in principle sufficient to retrieve the energy levels of all the states that belong to the ensemble~\cite{deur2019ground}. Providing a proper description of the true physical ensemble energy, through an appropriate {\it ensemble weight-dependent} Hxc density functional is, however, a very challenging task~\cite{yang2017direct,sagredo2018can,Cernatic2022,loos2020weight,marut2020weight,gould2021double,Yang2021_Second,gould2022single,gould2023local}.

As shown recently by the authors and co-workers~\cite{cernatic2024neutral}, the weight dependence of the ensemble Hxc density functional can be explicitly connected to the density-functional exactification of the one-electron KS picture (excitation energy-wise), through the formulation of exact Koopmans' theorems for specific ionization processes. Indeed, by combining the ionization of the ground state with that of the neutrally-excited state of interest, as originally proposed by Levy~\cite{levy1995excitation}, it becomes possible to exactify the KS orbital energies in the evaluation of neutral excitation energies. This can be achieved within the so-called extended $N$-centered (e$N$c) ensemble density-functional formalism~\cite{cernatic2024neutral}, where, by construction, the ensemble density still integrates to the {\it integer} number $N$ of electrons in the reference ground state, like in TGOK-DFT, despite the incorporation of charged excited states into the ensemble. This trick allows for an exactification of Koopmans' theorem without invoking the asymptotic behavior of the density away from the system under study, unlike in more conventional approaches to density-functional ensembles of ground and excited states~\cite{levy1995excitation,gould2022single}. An immediate consequence of such an exactification is the appearance of a density-functional derivative discontinuity in the Hxc potential following the inclusion of a given excitation into the ensemble. Even though e$N$c ensemble DFT is a very general approach, only single excitations from the highest occupied molecular orbital (HOMO) have been discussed in detail in Ref.~\citenum{cernatic2024neutral}. In the present work, we extend the discussion to any type of single or double excitation process, with a particular focus on the derivative discontinuity that the latter induces and the connection between the ensemble density-functional KS electronic structure and that of the true physical system.    

The paper is organized as follows.
After a brief review of e$N$c ensemble DFT in Sec.~\ref{sec:review_eNc_eDFT}, we present in Sec.~\ref{sec:exactification_Koopmans} a general density-functional exactification of Koopmans' theorem and its application to the evaluation of any single or double neutral excitation energy. The degree of excitation in the ensemble density-functional KS system and its connection to the physical process is also discussed (in Sec.~\ref{sec:general_blabla_true_and_KS_excitations}). A more explicit derivation of the theory for an e$N$c ensemble with two neutral excited states, in addition to the cationic ground state, is presented in Sec.~\ref{sec:general_construction_4-state}. Its exact implementation within the Hubbard dimer model is finally discussed (in Secs.\ref{sec:intro_HD_model} and \ref{sec:exact_pot_and_func_HD}), and the results obtained for various  ensemble weight values, correlation, and asymmetry regimes are analyzed in Sec.~\ref{sec:results_HD}. Conclusions are given in Sec.~\ref{sec5}.

\section{Theory}\label{sec:Theory}

\subsection{Brief review of extended $N$-centered ensemble DFT}\label{sec:review_eNc_eDFT}

While a regular $N$-centered ensemble consists of a reference $N$-electron ground state complemented by the cationic [$(N-1)$-electron] and anionic [$(N+1)$-electron] ground states, to which (possibly different) ensemble weights are assigned~\cite{senjean2018unified,senjean2020n}, an e$N$c ensemble incorporates neutral excitation processes~\cite{cernatic2024neutral}. In Ref.~\cite{cernatic2024neutral}, these processes have been considered explicitly for the $N$-electron system only but in fact, as it will become clear and useful in the following, excited states of the $(N\pm p)$-electron system, where $p=1,2,\ldots$, can be trivially incorporated into the ensemble too, thus making the formalism very general. Mathematically, an e$N$c ensemble, that we simply refer to as ensemble from now on, is described by the following density matrix operator,   
\begin{subequations}
\begin{align}
\Gammaop^{\bxi} &\overset{{\rm e}N{\rm c}}{=}
\left(1-\sum_{\nu\neq 0}\dfrac{N_{\nu}}{N}\xinu\right)\vert\Psi_0\rangle\langle\Psi_0\vert +
\sum_{\nu\neq 0}\xinu
\vert\Psi_{\nu}\rangle\langle\Psi_{\nu}\vert
\\
\label{eq:eNc_ens_shorthand_notation}
&\underset{\text{notation}}{\overset{\text{shorthand}}{=}}\sum_\nu\xi_\nu \vert\Psi_{\nu}\rangle\langle\Psi_{\nu}\vert
.
\end{align}
\end{subequations}
$\Psi_{0}\equiv\Psi_{0}^N$ denotes the (normalized) reference ground-state wavefunction of $N_0=N$ electrons with Hamiltonian $\hat{H}=\hat{T}+\hat{W}_{\rm ee}+\hat{V}_{\rm ext}$, where $\hat{T}\equiv -\frac{1}{2}\sum^N_{i=1}\nabla^2_{\br_i}$ is the kinetic energy operator, $\hat{W}_{\rm ee}\equiv\sum^N_{1\leq i<j}\frac{1}{\vert {\br_i-\br_j}\vert}\times$ is the electronic repulsion operator, and $\hat{V}_{\rm ext}=\int d{\br}\, v_{\rm ext}(\br)\,\hat{n}(\br)$ is the external potential (\ie, the nuclear attraction potential in conventional quantum chemistry computations) operator, $\hat{n}(\br)\equiv \sum^{N}_{i=1}\delta(\br -\br_i)\times$ being the electron density operator at position $\br$. $\left\{\Psi_{\nu}\right\}_{\nu\neq 0}$ are the remaining normalized $N_\nu$-electron eigenfunctions of $\hat{H}$ (which is now extended to the entire Fock space), with $N_\nu=N\pm p$ and $p=0,1,2,\ldots$, to which positive ensemble weights $\left\{\xi_\nu\right\}_{\nu\neq 0}$ are assigned. Note that the ensemble weight $\xi_0$, which is assigned to the reference $N$-electron ground state, is fully determined from the (charged or neutral) excited-state weights $\bxi\equiv\left\{\xi_\nu\right\}_{\nu\neq 0}$: 
\be
\xi_0\equiv \xi_0(\bxi)=1-\sum_{\nu\neq 0}\dfrac{N_{\nu}}{N}\xinu.
\ee
Most importantly, it ensures, by construction, that the ensemble electronic density
\be\label{eq:physical_eNc_dens}
\begin{split}
n^{\bxi}(\br):=
\Tr{\left[\Gammaop^{\bxi}\hat{n}(\br)\right]} 
&= 
\left(1-\sum_{\nu\neq 0}\dfrac{N_{\nu}}{N}\xinu\right)
n_{\Psi_0}(\br)
\\
&\quad+\sum_{\nu\neq 0}\xinu n_{\Psi_\nu}(\br),
\end{split}
\ee
where $\Tr$ denotes the trace,
integrates to the (so-called central) number $N$ of electrons in the reference ground state,
\be\label{eq:eNc_dens_dens_integrates_to_N}
\int\,\dr\, n^{\bxi}(\br) = \int\,\dr\, n_{\Psi_0}(\br)=N_0=N,
\ee
hence the name given to the ensemble. This particular constraint, which does not exist in the conventional Perdew--Parr--Levy--Balduz (PPLB) DFT of charged electronic excitations~\cite{perdew1982density,perdew1983physical} (see also Refs.~\citenum{baerends2013kohn,
baerends2017kohn,baerends2018density,Baerends2022_Chemical,Baerends2020_On_derivatives}), has fundamental implications that have been extensively discussed in previous works~\cite{Cernatic2022,cernatic2024neutral} and that will be exploited in the following, in particular for exactifying KS orbital energies in the evaluation of single- or multiple-electron neutral excitation energies.

In this context, the ensemble energy $E^{\bxi} := \Tr{\left[\Gammaop^{\bxi}\hat{H}\right]}$ reads
\be\label{eq:eNc_ener_explicit_exp}
E^{\bxi}=
\left(1-\sum_{\nu\neq 0}\dfrac{N_{\nu}}{N}\xinu\right)E_0 +
\sum_{\nu\neq 0}\xinu E_{\nu}.
\ee
When the ensemble weights $\left\{\xi_\nu\right\}_{N_\nu=N\pm p}$ assigned to all the states (including the reference $N$-electron ground state when $p=0$) belonging to a given $(N\pm p)$-electron sector ($p=0,1,2\ldots$) of the Fock space are monotonically decreasing with the energy, the ensemble energy can be determined variationally~\cite{cernatic2024neutral,gross1988rayleigh}, for {\it fixed} $\bxi$ weight values, as follows,    
\be
E^{\bxi}=\min_{\hat{\gamma}^{\bxi}}\Tr{\left[\hat{\gamma}^{\bxi}\hat{H}\right]},
\ee
where $\hat{\gamma}^{\bxi}$ is a trial ensemble density matrix operator.
The density functionalization of the theory emerges naturally from Levy's constrained search formalism~\cite{levy1979universal}, \ie,
\begin{subequations}
\begin{align}
&E^{\bxi}=\min_n \left\{
\min_{\hat{\gamma}^{\bxi}\rightarrow n}
\Tr{\left[\hat{\gamma}^{\bxi}\hat{H}\right]}
\right\}
\\
\label{eq:VP_Levy_eNc_ener}
&=\min_n \left\{
\min_{\hat{\gamma}^{\bxi}\rightarrow n}
\Tr{\left[\hat{\gamma}^{\bxi}\left(\hat{T}+\hat{W}_{\rm ee}\right)\right]}
+\int d\br\, v_{\rm ext}(\br)\,n(\br)
\right\},
\end{align}
\end{subequations}
where $n$ is a trial ensemble density and the density constraint $\hat{\gamma}^{\bxi}\rightarrow n$ reads
\be
n_{\hat{\gamma}^{\bxi}}(\br):=\Tr{\left[\hat{\gamma}^{\bxi}\hat{n}(\br)\right]}=n(\br).
\ee
On that basis, a general ensemble KS-DFT, where both neutral and charged electronic excitations are described, in principle exactly, can be formulated. Indeed, by rewriting Eq.~(\ref{eq:VP_Levy_eNc_ener}) as follows, 
\be
E^{\bxi}=\min_n \left\{
\min_{\hat{\gamma}^{\bxi}\rightarrow n}
\Tr{\left[\hat{\gamma}^{\bxi}\hat{T}\right]}
+E^{\bxi}_{\rm Hxc}[n]
+\int d\br\, v_{\rm ext}(\br)\,n(\br)
\right\},
\ee
where 
\begin{subequations}\label{eq:LL_def_ens_Hxc_func}
\begin{align}
E^{\bxi}_{\rm Hxc}[n]&=\min_{\hat{\gamma}^{\bxi}\rightarrow n}
\Tr{\left[\hat{\gamma}^{\bxi}\left(\hat{T}+\hat{W}_{\rm ee}\right)\right]}-\min_{\hat{\gamma}^{\bxi}\rightarrow n}
\Tr{\left[\hat{\gamma}^{\bxi}\hat{T}\right]}
\\
&:=F^{\bxi}[n]-T^{\bxi}_{\rm s}[n]
\end{align}
\end{subequations}
is the {\it weight-dependent} analogue for e$N$c ensembles of the Hxc density functional, we finally obtain the following exact variational expression of the ensemble energy,
\begin{subequations}
\begin{align}
\nonumber
E^{\bxi}&=\min_n \Bigg\{
\min_{\hat{\gamma}^{\bxi}\rightarrow n}
\Big\{
\Tr{\left[\hat{\gamma}^{\bxi}\hat{T}\right]}
+E^{\bxi}_{\rm Hxc}[n_{\hat{\gamma}^{\bxi}}]
\\
&\quad\quad\quad\quad+\int d\br\, v_{\rm ext}(\br)\,n_{\hat{\gamma}^{\bxi}}(\br)
\Big\}
\Bigg\}
\\
\label{eq:VP_gamma_eNc_KS-DFT}
&=
\min_{\hat{\gamma}^{\bxi}}
\left\{
\Tr{\left[\hat{\gamma}^{\bxi}\left(\hat{T}+\hat{V}_{\rm ext}\right)\right]}
+E^{\bxi}_{\rm Hxc}[n_{\hat{\gamma}^{\bxi}}]
\right\}.
\end{align}
\end{subequations}
The minimizing ensemble in Eq.~(\ref{eq:VP_gamma_eNc_KS-DFT}) consists of (weight-dependent) noninteracting KS $N_\nu$-electron wavefunctions 
$\Phi^{\bxi}_\nu$
(\ie, Slater determinants or configuration state functions) that reproduce the true physical ensemble density of Eq.~(\ref{eq:physical_eNc_dens}),  
\be
\left(1-\sum_{\nu\neq 0}\dfrac{N_{\nu}}{N}\xinu\right)
n_{\Phi^{\bxi}_0}(\br)
+\sum_{\nu\neq 0}\xinu n_{\Phi^{\bxi}_\nu}(\br)=n^{\bxi}(\br).
\ee
They fulfill the following self-consistent equation,
\be\label{eq:ens_KS_eqs_many-body_KS_states}
\left[\hat{T}+\hat{V}_{\rm
ext}+\int d\br\,v^{\bxi}_{\rm
Hxc}(\br)\,\hat{n}(\br)\right]\myket{\Phi^{\bxi}_\nu}=\mathcal{E}^{\bxi}_\nu\myket{\Phi^{\bxi}_\nu},\; \forall\nu,
\ee
where $v^{\bxi}_{\rm
Hxc}(\br)=\left.\delta E_{\rm Hxc}^{\bxi}[n]/\delta
n(\br)\right|_{n=n^{\bxi}}$ is the ensemble Hxc potential. Solving Eq.~(\ref{eq:ens_KS_eqs_many-body_KS_states}) is equivalent to solving the self-consistent one-electron-like ensemble KS equations,
\be\label{eq:ens_KS_eqs_orbitals}
\left[-\dfrac{\nabla_{\br}^2}{2}+v_{\rm ext}(\br)+v^{\bxi}_{\rm
Hxc}(\br)\right]\varphi^{\bxi}_k(\br)=\varepsilon^{\bxi}_k\varphi^{\bxi}_k(\br),
\ee
from which the ensemble density and the total (fictitious) KS energies can be determined. Indeed, if we denote $n_{\nu,k}$ the integer occupation of the KS orbital $\varphi^{\bxi}_k(\br)$ in the KS state $\Phi^{\bxi}_\nu$ (note that $\sum_k n_{\nu,k}=N_\nu$) and we use the shorthand notation of Eq.~(\ref{eq:eNc_ens_shorthand_notation}), then  
\be
\begin{split}
n^{\bxi}(\br)&=\sum_\nu\xi_\nu\sum_k n_{\nu,k}\left\vert\varphi^{\bxi}_k(\br)\right\vert^2
\\
&=\sum_k \left(\sum_\nu\xi_\nu n_{\nu,k}\right)\left\vert\varphi^{\bxi}_k(\br)\right\vert^2,
\end{split}
\ee
where, as readily seen, the fractional occupations of the KS orbitals are controlled by the ensemble weights,
and
\be\label{eq:total_KS_ener_from_KS_orb_ener}
\mathcal{E}^{\bxi}_\nu=\sum_k n_{\nu,k}\,\varepsilon^{\bxi}_k.
\ee
As readily seen from Eq.~(\ref{eq:ens_KS_eqs_orbitals}), the analog for ensembles of the KS potential is simply obtained by adding to the physical external potential the (ensemble) Hxc potential, like in a regular DFT calculation:
\be\label{eq:ens_KS_potential_def}
v^{\bxi}_{\rm KS}(\br)=v_{\rm ext}(\br)+v^{\bxi}_{\rm
Hxc}(\br).
\ee

The e$N$c ensemble energy introduced in Eq.~(\ref{eq:eNc_ener_explicit_exp}) is an auxiliary quantity which has {\it a priori} no physical meaning. Its evaluation as a function of the ensemble weights $\bxi$ is, however, of high interest. Indeed, the fact that it varies linearly with $\bxi$ enables to extract any ground- or excited-state energy level as follows, 
\begin{subequations}
\begin{align}
E_\nu & \underset{\nu\geq 0}{=}\dfrac{N_\nu}{N}E_0+\sum_{\lambda\neq 0}\delta_{\lambda\nu}\left(E_\lambda-\dfrac{N_\lambda}{N}E_0\right)
\\
&=\dfrac{N_\nu}{N}\left(E^{\bxi}-\sum_{\lambda\neq 0}\xi_\lambda\dfrac{\partial E^{\bxi}}{\partial \xi_\lambda}\right)+\sum_{\lambda\neq 0}\delta_{\lambda\nu}\dfrac{\partial E^{\bxi}}{\partial \xi_\lambda}
\\
\label{eq:ener_level_from_eNc_ener}
&=\dfrac{N_\nu}{N}E^{\bxi}
+\sum_{\lambda\neq 0}\left(\delta_{\lambda\nu}-\dfrac{N_\nu}{N}\xi_{\lambda}\right)\dfrac{\partial E^{\bxi}}{\partial \xi_{\lambda}}
,
\end{align}
\end{subequations}
and therefore any (neutral or charged) excitation energy, by difference. From Eq.~(\ref{eq:ener_level_from_eNc_ener}) and the variational KS-DFT expression of the ensemble energy in Eq.~(\ref{eq:VP_gamma_eNc_KS-DFT}) we can finally evaluate any physical $\nu\rightarrow\kappa$ excitation energy from the KS one, in principle exactly, as follows~\cite{cernatic2024neutral},  
\be\label{eq:Excitation-energies_Exact-vs-KS-ones}                   
\begin{aligned}                                                        
&E_\kappa-E_\nu=\mathcal{E}^{\bxi}_\kappa-\mathcal{E}^{\bxi}_\nu
\\                                                                  
&                                                       
+\dfrac{\left(N_\kappa-N_\nu\right)}{N}\left(E_{\rm Hxc}^{\bxi}[n^{\bxi}]-\int d\br\, v_{\rm Hxc}^{\bxi}(\br)n^{\bxi}(\br)\right)
\\                                          
&                                       
+                                       
\sum_{\lambda\neq 0}\left(                   
\delta_{\lambda\kappa}-\delta_{\lambda\nu}-\dfrac{\left(N_\kappa-N_\nu\right)}{N}\xilambda\right)\left.\dxilambda{E_{\rm
Hxc}^{\bxi}[n]}\right|_{n=n^{\bxi}}. 
\end{aligned}                            
\ee
While Eq.~(\ref{eq:Excitation-energies_Exact-vs-KS-ones}) has been exploited in Ref.~\citenum{cernatic2024neutral} to exactify KS orbital energies in the description of single-electron excitations from the HOMO to higher KS orbitals, we will consider in the following more general excitation processes, including double excitations.

\subsection{Exactification of KS orbital energies for single- and multiple-electron excitations}\label{sec:exactification_Koopmans}

A key feature of the e$N$c ensemble formalism is that, even when we describe charged excitation processes, Eq.~(\ref{eq:Excitation-energies_Exact-vs-KS-ones}) remains {\it invariant} under any constant shift $v_{\rm Hxc}^{\bxi}(\br)\rightarrow v_{\rm Hxc}^{\bxi}(\br)+c$ in the ensemble Hxc potential (see Eq.~(\ref{eq:eNc_dens_dens_integrates_to_N})):
\be
\begin{split}
&\mathcal{E}^{\bxi}_\kappa-\mathcal{E}^{\bxi}_\nu
-\dfrac{\left(N_\kappa-N_\nu\right)}{N}\int d\br\, v_{\rm Hxc}^{\bxi}(\br)n^{\bxi}(\br)
\\
&=\left(\mathcal{E}^{\bxi}_\kappa+N_\kappa c\right)-\left(\mathcal{E}^{\bxi}_\nu+N_\nu c\right)
\\
&\quad \quad-\dfrac{\left(N_\kappa-N_\nu\right)}{N}\int d\br\, \left(v_{\rm Hxc}^{\bxi}(\br)+c\right)n^{\bxi}(\br)
.
\end{split}
\ee
This degree of freedom in the theory allows for a systematic exactification of Koopmans' theorem~\cite{cernatic2024neutral}, as further explained in the following. Let us consider, for example, the {\it single-electron} ionization process $\nu\rightarrow \kappa$ of an $N$-electron ground or excited state $\nu\geq 0$ (\ie, $N_\nu=N$ and $N_\kappa=N-1$), where, unlike in Ref.~\citenum{cernatic2024neutral}, $\kappa$ can be an excited $(N-1)$-electron state. According to Eq.~(\ref{eq:Excitation-energies_Exact-vs-KS-ones}), the KS ionization energy matches the physical one, \ie~(see Eq.~(\ref{eq:total_KS_ener_from_KS_orb_ener})),
\be
E_\kappa-E_\nu=\mathcal{E}^{\bxi}_\kappa-\mathcal{E}^{\bxi}_\nu=\sum_k\left(n_{\kappa,k}-n_{\nu,k}\right)\varepsilon^{\bxi}_k,
\ee
if and only if

\be\label{eq:general_constraint_Koopmans_Hcv_pot}
\int \dfrac{d\br}{N}\, v_{\rm Hxc}^{\bxi}(\br)n^{\bxi}(\br)
\underset{N_\nu-N_\kappa=1}{\overset{\nu\rightarrow \kappa}{=}}
\mathscr{D}^{{\bxi} [\nu\rightarrow \kappa]}_{\rm Hxc}[n^{\bxi}],
\ee
where
\be\label{eq:constant_defining_pot_uniquely}
\mathscr{D}^{{\bxi} [\nu\rightarrow \kappa]}_{\rm Hxc}[n]:=\dfrac{E_{\rm Hxc}^{\bxi}[n]}{N}
+\sum_{\lambda\neq 0}\left(\delta_{\lambda\nu}                   
-\delta_{\lambda\kappa}-\dfrac{\xilambda}{N}\right)\dxilambda{E_{\rm
Hxc}^{\bxi}[n]}.
\ee
Note that Eq.~(\ref{eq:general_constraint_Koopmans_Hcv_pot}) defines the Hxc potential uniquely, not up to a constant anymore. We denote the latter potential $v_{\rm Hxc}^{\bxi [\nu\rightarrow \kappa]}(\br)$ in the following.\\

On that basis, we can express any neutral excitation energy ${\Omega^N_\nu=E^N_\nu-E^N_0}$ in terms of the KS orbital energies, simply by considering two {\it distinct} ionization processes, namely the ionization $[0\rightarrow \kappa]$ of the $N$-electron ground state and the ionization $[\nu\rightarrow \kappa]$ of the $N$-electron excited state $\nu>0$ of interest:
\begin{subequations}
\begin{align}
&\Omega^N_\nu\equiv E_\nu-E_0=\left(E_\kappa-E_0\right)-\left(E_\kappa-E_\nu\right)
\\
&=\left(\mathcal{E}^{\bxi[0\rightarrow \kappa]}_\kappa-\mathcal{E}^{\bxi[0\rightarrow \kappa]}_0\right)-\left(\mathcal{E}^{\bxi[\nu\rightarrow \kappa]}_\kappa-\mathcal{E}^{\bxi[\nu\rightarrow \kappa]}_\nu\right)
\\
\label{eq:XE_from_KS_orb_energies_and_occ}
&=\sum_k\left(n_{\kappa,k}-n_{0,k}\right)\varepsilon^{\bxi[0\rightarrow \kappa]}_k-\sum_k\left(n_{\kappa,k}-n_{\nu,k}\right)\varepsilon^{\bxi[\nu\rightarrow \kappa]}_k.
\end{align}
\end{subequations}
Interestingly, in the above mathematical construction, the Hxc potentials associated with each ionization process reproduce the same ensemble density $n^{\bxi}(\br)$. Consequently, they differ by a constant which, according to Eqs.~(\ref{eq:general_constraint_Koopmans_Hcv_pot}) and (\ref{eq:constant_defining_pot_uniquely}), simply corresponds to a weight derivative of the Hxc ensemble density functional: 
\be\label{eq:DD_general_case}
\int \dfrac{d\br}{N}\, \left(v_{\rm Hxc}^{\bxi[\nu\rightarrow \kappa]}(\br)-v_{\rm Hxc}^{\bxi[0\rightarrow \kappa]}(\br)\right)n^{\bxi}(\br)
{\overset{\nu>0}=}\left.\dfrac{\partial E_{\rm
Hxc}^{\bxi}[n]}{\partial \xi_\nu}\right|_{n=n^{\bxi}}.
\ee
Eq.~(\ref{eq:DD_general_case}) generalizes previous work~\cite{levy1995excitation,gould2022single} to any type of neutral excitation, without invoking the asymptotic behavior of the ensemble density away from the system of interest (see Refs.~\citenum{PRA21_Hodgson_exact_Nc-eDFT_1D} and \citenum{Cernatic2022} for a detailed comparison of the two approaches for charged excitations). 

If $\kappa$ corresponds, in the noninteracting KS world (see Secs.~\ref{sec:general_blabla_true_and_KS_excitations} and \ref{sec:analysis_WFs_KS_rep_HF} for further discussion on this point), to an ionized state with a hole in the KS orbital $i$ ($1\leq i\leq N$) while $\nu>0$ corresponds to a single $i\rightarrow a$ excitation ($a > N$), then the corresponding exact physical excitation energy simply reads, according to Eq.~(\ref{eq:XE_from_KS_orb_energies_and_occ}),     
\be\label{eq:exact_general_single_excitation_ener}
\Omega^N_\nu=\varepsilon^{\bxi[\nu\rightarrow \kappa]}_a-\varepsilon^{\bxi[0\rightarrow \kappa]}_i.
\ee

On the other hand, if $\nu>0$ now corresponds (still in the KS world) to a double $(i,j)\rightarrow (a,b)$ excitation ($1\leq j\leq N$ and $b > N$), and $\kappa$ is still the singly-ionized state with a hole in orbital $i$, then the exact physical excitation energy expression becomes, according to Eq.~(\ref{eq:XE_from_KS_orb_energies_and_occ}),
\begin{subequations}\label{eq:exact_general_doubles_ener}
\begin{align}
\Omega^N_\nu&=-\varepsilon^{\bxi[0\rightarrow \kappa]}_i-\varepsilon^{\bxi[\nu\rightarrow \kappa]}_j+\varepsilon^{\bxi[\nu\rightarrow \kappa]}_a+\varepsilon^{\bxi[\nu\rightarrow \kappa]}_b
\\
&=\varepsilon^{\bxi[\nu\rightarrow \kappa]}_a-\varepsilon^{\bxi[0\rightarrow \kappa]}_i+\left(\varepsilon^{\bxi[\nu\rightarrow \kappa]}_b-\varepsilon^{\bxi[\nu\rightarrow \kappa]}_j\right)
,
\end{align}
\end{subequations}
or, equivalently, 
\be\label{eq:exact_general_doubles_ener2}
\Omega^N_\nu=\varepsilon^{\bxi[\nu\rightarrow \kappa]}_a-\varepsilon^{\bxi[0\rightarrow \kappa]}_i+\left(\varepsilon^{\bxi[0\rightarrow \kappa]}_b-\varepsilon^{\bxi[0\rightarrow \kappa]}_j\right),
\ee
because the Hxc potentials $v_{\rm Hxc}^{\bxi [\nu\rightarrow \kappa]}(\br)$ and $v_{\rm Hxc}^{\bxi [0\rightarrow \kappa]}(\br)$ only differ by a constant expressed in Eq.~(\ref{eq:DD_general_case}).
Eqs.~(\ref{eq:exact_general_single_excitation_ener}) and (\ref{eq:exact_general_doubles_ener2}) provide an exactification of the KS orbital energies in the evaluation of single- and double-electron excitation energies, respectively. They generalize Eq.~(54) of Ref.~\citenum{cernatic2024neutral} which is only applicable to single excitations from the HOMO.  

\subsection{What are we supposed to learn from the KS ensemble about physical excitation processes?}\label{sec:general_blabla_true_and_KS_excitations}

As already mentioned in the introduction, the description of double electronic excitations (\ie, the modelling of two-hole/two-particle states) in the context of linear response TD-DFT is very challenging~\cite{Huix-Rotllant2011_Assessment,Lacombe2023_Non-adiabatic}. Indeed, in the latter regime, only single excitations (\ie, one-hole/one-particle states) are treated {\it explicitly}. Double electron excitation energies can in principle be retrieved by using a proper frequency-dependent Hxc kernel~\cite{Huix-Rotllant2011_Assessment,Lacombe2023_Non-adiabatic}. The situation is quite different in the context of ensemble DFT, since multiple electronic excitations can be explicitly incorporated into the KS ensemble. What is far from clear, however, is how informative the KS ground and excited states are about the true interacting eigenstates. 
Let us first comment on a common misunderstanding of the statement ``ensemble DFT can describe double excitations". Obviously, the latter does not mean that the true physical excitation process (to which double excitations may contribute) matches the one occuring in the ensemble density-functional KS system. It simply means that two-hole/two-particle excitation processes can be treated explicitly within the ensemble KS orbital space. Despite the loss of information about the true interacting states, which is a common feature of density-functional theories, ensemble DFT still provides an in-principle exact description of single and multiple excitations, ensemble density-wise. Indeed, for a given number $\mathscr{M}$ of lowest $N$-electron states ($\mathscr{M}=3$ in the following) and given ensemble weight values, the noninteracting KS ensemble, which contains the same number $\mathscr{M}$ of lowest $N$-electron KS eigenstates (Slater determinants or configuration state functions) as the physical one, is expected to reproduce the true interacting ensemble density. It is {\it a priori} its only connection with the true physical ensemble but it is sufficient to determine, in principle exactly, the energy levels of all the states that belong to that ensemble, according to Eqs.~(\ref{eq:VP_gamma_eNc_KS-DFT}) and (\ref{eq:ener_level_from_eNc_ener}) [see also Refs.~\citenum{deur2019ground} and \citenum{cernatic2024neutral}]. The identification of excitations is clear in the noninteracting KS picture. For example, in the Hubbard dimer model (see Sec.~\ref{sec:Application}), the first excited state is singly-excited and the second one is doubly-excited. However, true interacting electronic structures are much more complex. They can be mixtures of ground, singly-excited, and doubly-excited KS states, for example. In some specific asymmetry and correlation regimes, a reordering of the eigenstates may also occur when switching from the noninteracting ensemble KS picture to the interacting one. These different scenarios are illustrated and further discussed in Sec.~\ref{sec:analysis_WFs_KS_rep_HF}.

\section{Explicit formulation involving the ground cationic state and two neutral excited states}\label{sec:general_construction_4-state}

We consider in this section the particular case (studied later in the Hubbard dimer model) of an e$N$c ensemble consisting of the reference $N$-electron ground state, the two lowest $N$-electron excited states (with weights $\xi_1$ and $\xi_2$, respectively), and the $(N-1)$-electron ground state (with weight $\xi_-$):
\be
\begin{split}
\Gammaop^{\bxi}& = 
\left(1-\dfrac{(N-1)}{N}\xi_{-} - \xi_1 - \xi_2 \right)\ketbra{\Psi_0^N} 
\\
&+\xi_{-}\ketbra{\Psi_0^{N-1}} + \xi_1 \ketbra{\Psi_1^{N}} + \xi_2 \ketbra{\Psi_2^{N}},  
\end{split}
\ee
where the collection of independent weights reduces to
\be
{\bxi \equiv (\xi_-,\xi_1,\xi_2)}.
\ee
Note that, in order to allow for a variational evaluation of the corresponding ensemble energy,
\be\label{eq:eNc_ens_ener_4states}
\begin{split}
E^{\bxi} &= \left(1-\dfrac{(N-1)\xi_-}{N} - \xi_1 -\xi_2\right)E_0^N + \xi_- E_0^{N-1}
\\
&\quad+ \xi_1 E_1^N + \xi_2 E_2^N,
\end{split}
\ee
which is necessary to set up an ensemble DFT, the following inequalities should be fulfilled: 
\be
\xi_-\geq 0
\ee
and~\cite{gross1988rayleigh}
\be\label{eq:eNc-Hdim-ineqs-expr1}
\xi_0=1-\dfrac{(N-1)\xi_-}{N} - \xi_1 -\xi_2 \geq \xi_1 \geq \xi_2\geq 0.
\ee
Consequently, we have 
\be
2\xi_2\leq 1-\dfrac{(N-1)\xi_-}{N} - \xi_1\leq 1-\dfrac{(N-1)\xi_-}{N}-\xi_2, 
\ee
thus leading to the following allowed range of ensemble weight values, 
\be\label{eq:range_xi2}
0\leq \xi_2\leq \dfrac{1}{3}\left(1-\dfrac{(N-1)\xi_-}{N}\right)  
\ee
and
\be\label{eq:range_xi1}
\xi_2\leq \xi_1\leq \dfrac{1}{2}\left(1-\dfrac{(N-1)\xi_-}{N}-\xi_2\right). 
\ee
Turning to the general construction in Eq.~(\ref{eq:general_constraint_Koopmans_Hcv_pot}) of the (unique) Hxc potential that satisfies Koopmans' theorem exactly, for a given ionization process that will be indexed with $\mathcal{I}=0,1,2$ in the following, we obtain from Eq.~(\ref{eq:constant_defining_pot_uniquely}) the more explicit expressions
\be\label{eq:Dfunc_ionization_GS}
\begin{split}
\mathscr{D}^{{\bxi} [0]}_{\rm Hxc}[n]&=\dfrac{E_{\rm Hxc}^{\bxi}[n]}{N}-\left(1+\dfrac{\xi_-}{N}\right)\dxim{E_{\rm Hxc}^{\bxi}[n]}-\dfrac{\xi_1}{N}\dxinum{{E_{\rm Hxc}^{\bxi}[n]}}{1}
\\
&\quad -\dfrac{\xi_2}{N}\dxinum{{E_{\rm Hxc}^{\bxi}[n]}}{2},
\end{split}
\ee
\be\label{eq:Dfunc_ionization_1stexcitation}
\begin{split}
\mathscr{D}^{{\bxi} [1]}_{\rm Hxc}[n]&=\dfrac{E_{\rm Hxc}^{\bxi}[n]}{N}-\left(1+\dfrac{\xi_-}{N}\right)\dxim{E_{\rm Hxc}^{\bxi}[n]}
\\
&\quad 
+\left(1-\dfrac{\xi_1}{N}\right)\dxinum{{E_{\rm Hxc}^{\bxi}[n]}}{1}
-\dfrac{\xi_2}{N}\dxinum{{E_{\rm Hxc}^{\bxi}[n]}}{2}
,
\end{split}
\ee
and
\be\label{eq:Dfunc_ionization_2ndexcitation}
\begin{split}
\mathscr{D}^{{\bxi} [2]}_{\rm Hxc}[n]&=\dfrac{E_{\rm Hxc}^{\bxi}[n]}{N}-\left(1+\dfrac{\xi_-}{N}\right)\dxim{E_{\rm Hxc}^{\bxi}[n]}-\dfrac{\xi_1}{N}\dxinum{{E_{\rm Hxc}^{\bxi}[n]}}{1}
\\
&\quad +\left(1-\dfrac{\xi_2}{N}\right)\dxinum{{E_{\rm Hxc}^{\bxi}[n]}}{2}
,
\end{split}
\ee
for the ionization of the ground state ($\mathcal{I}=0)$, the ionization of the first excited state ($\mathcal{I}=1$), and the ionization of the second excited state ($\mathcal{I}=2$), respectively. An exact implementation of the three Hxc potentials from the above ensemble density-functional quantities is presented in the next section within the Hubbard dimer model, as a proof of concept. 


\section{Exact implementation for the two-electron Hubbard dimer}\label{sec:Application}

\subsection{Introduction to the model}\label{sec:intro_HD_model}

The Hubbard dimer is a simple but nontrivial two-site lattice model that can be used, for example, for describing diatomic molecules~\cite{li2018density}. As it can be solved exactly~\cite{carrascal2015hubbard}, it is often used as a toy system for testing new ideas in connection with the many-body problem~\cite{carrascal2015hubbard,li2018density,deur2018exploring,
sagredo2018can,carrascal2018linear,smith2016exact,deur2019ground,
Cernatic2022,Giarrusso2023_Exact,Ullrich2018_Density,scott2023exact,Sobrino2023_What,Liebert2023_Refining}. The basic idea of the model is to simplify the (second-quantized) {\it ab initio} Hamiltonian as follows,
\be\label{eq:Hamiltonian_def}
\hat{H}\rightarrow \hat{\mathcal{H}}=\hat{\mathcal{T}}+\hat{\mathcal{U}}+\hat{\mathcal{V}}_{\rm
ext},
\ee
where the analogue for the kinetic energy operator $\hat{\mathcal{T}}$ (the so-called hopping operator), the on-site electron repulsion operator $\hat{\mathcal{U}}$, and the local
(external) potential operator $\hat{\mathcal{V}}_{\rm
ext}$ read
\begin{subequations}\label{eq:Hamiltonian_Hdim}
\begin{align}
\hat{\mathcal{T}}&=-t\sum_{\sigma = \uparrow,\downarrow}
(\hat{c}^{\dagger}_{0\sigma}\hat{c}_{1\sigma} + \hat{c}^{\dagger}_{1\sigma}\hat{c}_{0\sigma}),
\label{eq:Hamiltonian_Hdim_hopping}
\\
\hat{\mathcal{U}}&=U\sum_{i=0}^{1}\hat{n}_{i\uparrow}\hat{n}_{i\downarrow},
\label{eq:Hamiltonian_Hdim_U}
\\
\hat{\mathcal{V}}_{\rm ext}&=\frac{\Delta v_{\rm ext}}{2}(\hat{n}_{1} - \hat{n}_{0}),
\label{eq:Hamiltonian_Hdim_vext}
\end{align}
\end{subequations}
respectively. The index $i \in \left\{0,1\right\}$ labels the two atomic sites,  $\hat{n}_{i\sigma} = \hat{c}^{\dagger}_{i\sigma}\hat{c}_{i\sigma}$
is the spin-site occupation operator, and $\hat{n}_{i} = \sum_{\sigma =
\uparrow,\downarrow}\hat{n}_{i\sigma}$ plays the role of
the density operator (on site $i$). The asymmetry of the model is controlled by the difference $\Delta
v_{\rm ext}$ in external potential between sites 1 and 0, while electron correlation effects can be tuned through the ratio $U/t$. In this context, the electron density is the collection of site occupations $\left\{n_i=\langle\hat{n}_i\rangle\right\}_{i=0,1}$. In the following, the central number of electrons will be fixed to $N=n_0+n_1=2$, so that the density reduces to a single number $n$ that we choose to be the occupation of site 0, \ie, $n:=n_0$. Note that, in the symmetric dimer (which would correspond to the hydrogen molecule in a minimal basis, for example), we have $n=1$. The asymmetric dimer can be used, on the other hand, as a model for heteronuclear diatomic molecules such as LiF~\cite{li2018density}, for example. 

We consider in the following the e$N$c ensemble described in Sec.~\ref{sec:general_construction_4-state}, where the two neutral (singlet) excited states are, in the noninteracting KS picture, singly and doubly excited, respectively.  
The hopping parameter is set to $t=1/2$ throughout the paper. 


\subsection{Computation of exact ensemble density-functional energies and potentials}\label{sec:exact_pot_and_func_HD} 

The implementation of e$N$c ensemble DFT for a tri-ensemble (\ie, in the particular case where $\xi_2=0$) has been extensively discussed in Ref.~\citenum{cernatic2024neutral}. As shown in Appendix~\ref{app1}, the more general 4-state ensemble case studied in the present work can be recast into an effective tri-ensemble problem, simply because the three two-electron ground- and excited-state (singlet) energies sum up to $2U$~\cite{deur2019ground}. This simplification, which applies to the Hubbard dimer only and is not general, leads to the following expression for the interacting ensemble density functional introduced in Eq.~(\ref{eq:LL_def_ens_Hxc_func}), 
\be\label{eq:4_to_3-state_ens_Lieb_func_HD}
F^{\bxi}(n) = 2U\xi_2 + (1-3\xi_2)F^{\bzeta}(\nu),
\ee
where $\bzeta = (\zeta_-,\zeta_1)$ is an effective tri-ensemble weights collection defined as follows,
\begin{subequations}\label{eq:effective_weights_HD}
\begin{align}
\zeta_- &\equiv \zeta_-({\bxi})=\dfrac{\xi_-}{1-3\xi_2},
\\
\zeta_1 &\equiv \zeta_1({\bxi})= \dfrac{\xi_1 - \xi_2}{1-3\xi_2},
\end{align}
\end{subequations}
and 
\be\label{eq:effective_dens_HD}
\nu \equiv \nu(n,{\bxi})= \dfrac{n-3\xi_2}{1-3\xi_2}
\ee
is an effective tri-ensemble density. From Eq.~(\ref{eq:4_to_3-state_ens_Lieb_func_HD}), taken at $U=0$, which gives
\be\label{eq:4_to_3-state_ens_Ts_func_HD}
T_{\rm s}^{\bxi}(n)=(1-3\xi_2)T_{\rm s}^{\bzeta}(\nu),
\ee
and the following expression for the tri-ensemble noninteracting kinetic energy functional~\cite{cernatic2024neutral}, 
\be
T_{\rm s}^{\bzeta}(\nu)=-2t\sqrt{(1-\zeta_1)^2 - (1-\nu)^2},
\ee
we can express exactly and analytically the 4-state ensemble density-functional noninteracting kinetic energy as follows,
\be\label{eq:Ts_4-state_func_HD}
T_{\rm s}^{\bxi}(n) =
-2t\sqrt{(1-\xi_1-2\xi_2)^2 - (1-n)^2}.
\ee
Note that $T_{\rm s}^{\bxi}(n)$ does not depend on $\xi_-$~\cite{cernatic2024neutral}. Moreover, according to Eqs.~(\ref{eq:4_to_3-state_ens_Lieb_func_HD}) and (\ref{eq:4_to_3-state_ens_Ts_func_HD}), the 4-state ensemble density-functional Hxc energy can be evaluated from the tri-ensemble one (which can be computed exactly through a Lieb maximization~\cite{cernatic2024neutral}) as follows,
\begin{subequations}
\begin{align}
E^{\bxi}_{\rm Hxc}(n)&=F^{\bxi}(n)-T_{\rm s}^{\bxi}(n)
\\
\label{eq:Hxc_func_from_triens_HD}
&=2U\xi_2+(1-3\xi_2)E^{\bzeta}_{\rm Hxc}(\nu).
\end{align}
\end{subequations}
Turning to the ensemble density-functional potentials, the difference in KS potential between sites 1 and 0, $\Delta v_{\rm KS}^{\bxi}(n)$, is the maximizer~\cite{deur2017exact} for $U=0$ of the e$N$c ensemble Lieb functional introduced in Appendix~\ref{app1} (see Eq.~(\ref{eq:4-state_Lieb_func_HD})), thus leading to 
\be\label{eq:dens_func_KS_pot_from_deriv_HD}
\Delta v_{\rm KS}^{\bxi}(n)=\dfrac{\partial T_{\rm s}^{\bxi}(n)}{\partial n},
\ee
or, equivalently (see Eq.~(\ref{eq:Ts_4-state_func_HD})),
\be\label{eq:general_KS_pot_diff_HD}
\Delta v_{\rm KS}^{\bxi}(n) =
\dfrac{2t(n-1)}{\sqrt{(1-\xi_1 - 2\xi_2)^2 - (1-n)^2}}.
\ee
On the other hand, the ensemble Hxc potential {difference} can be evaluated as follows (see Eq.~(\ref{eq:ens_KS_potential_def})),
\be\label{eq:Hxc_pot_eq_KS_minus_ext_HD}
\Delta v_{\rm Hxc}^{\bxi}= \Delta v_{\rm KS}^{\bxi}(n^{\bxi}) - \Delta v_{\rm ext},
\ee
where $n^{\bxi}$ is the true physical ensemble density. The latter can be determined, for given ensemble weights $\bxi$ and external potential difference $\Delta v_{\rm ext}$ values, from the Hellmann--Feynman theorem~\cite{deur2017exact}:
\be
1-n^{\bxi}=\left.\dfrac{\partial E^{\bxi}(\Delta v)}{\partial \Delta v}\right|_{\Delta v=\Delta v_{\rm ext}}.
\ee
Note that $\Delta v_{\rm ext}$ is the maximizing potential of the interacting ensemble Lieb functional in Eq.~(\ref{eq:4-state_Lieb_func_HD}), when evaluated for $n=n^{\bxi}$. Consequently,  
\be
\Delta v_{\rm ext}=\left.\dfrac{\partial F^{\bxi}(n)}{\partial n}\right|_{n=n^{\bxi}},
\ee
and, according to Eqs.~(\ref{eq:dens_func_KS_pot_from_deriv_HD}) and (\ref{eq:Hxc_pot_eq_KS_minus_ext_HD}),  
\be
\Delta v_{\rm Hxc}^{\bxi}=\Delta v_{\rm Hxc}^{\bxi}(n^{\bxi}),
\ee
where the Hxc ensemble density-functional potential {\it difference}  formally equals 
\be\label{eq:Hxc_pot_-minus_deriv_n_HD} 
\Delta v_{\rm Hxc}^{\bxi}(n)=\dfrac{\partial T_{\rm s}^{\bxi}(n)}{\partial n}-\dfrac{\partial F^{\bxi}(n)}{\partial n}=-\dfrac{\partial E_{\rm Hxc}^{\bxi}(n)}{\partial n}.
\ee
Note that the minus sign on the right-hand side of the above equation originates from the arbitrary choice we made to compute potential differences between sites 1 and 0 while referring to the occupation of site 0 as the density $n$~\cite{senjean2017local,deur2017exact}.

We can now construct from $\Delta v_{\rm Hxc}^{\bxi}$ the ensemble Hxc potential, which is in principle defined up to a constant that we denote $- \mu^{\bxi}_{\rm Hxc}$. Its value on site $i$ ($i=0,1$) reads (see Eq.~(\ref{eq:Hamiltonian_Hdim_vext}))  
\be\label{eq:Hxc_potent_with_constant_HD}
v^{\bxi}_{{\rm Hxc},i} = (-1)^{i-1}\dfrac{\Delta v_{\rm Hxc}^{\bxi}(n^{\bxi})}{2} - \mu^{\bxi}_{\rm Hxc}.
\ee
For a given ionization process $\mathcal{I}$ (see Sec.~\ref{sec:general_construction_4-state}), the constant $- \mu^{\bxi}_{\rm Hxc}$ is uniquely defined from the constraint of Eq.~(\ref{eq:general_constraint_Koopmans_Hcv_pot}), which becomes in the Hubbard dimer model (we recall that $N=2$),
\be
\begin{split}
&\int \dfrac{d\br}{N}\, v_{\rm Hxc}^{\bxi[\mathcal{I}]}(\br)n^{\bxi}(\br)
\\
&\rightarrow \dfrac{1}{2}\sum^1_{i=0}v^{\bxi[\mathcal{I}]}_{{\rm Hxc},i}\,n^{\bxi}_i=\dfrac{\left(1-n^{\bxi}\right)\Delta v_{\rm Hxc}^{\bxi}(n^{\bxi})}{2}-\mu^{\bxi[\mathcal{I}]}_{\rm Hxc}
\\
&\hspace{2.7cm}\overset{!}{=}\mathscr{D}^{\bxi[\mathcal{I}]}_{\rm Hxc}(n^{\bxi})
,
\end{split}
\ee
thus ensuring the exactification of Koopmans' theorem for that specific ionization. We finally conclude from Eq.~(\ref{eq:Hxc_potent_with_constant_HD}) that the value of the corresponding Hxc potential on site 1 equals 
\be\label{eq:Hxc_pot_with_constant_unique_HD}
v^{\bxi[\mathcal{I}]}_{{\rm Hxc}}:=v^{\bxi[\mathcal{I}]}_{{\rm Hxc},1}{=}
\dfrac{n^{\bxi}\Delta v_{\rm Hxc}^{\bxi}(n^{\bxi})}{2}+\mathscr{D}^{\bxi[\mathcal{I}]}_{\rm Hxc}(n^{\bxi}).
\ee


\subsection{Results and discussion}\label{sec:results_HD}

\subsubsection{Derivative discontinuities induced by double excitations}\label{sec:DD_doubles_HD}

\begin{figure*}[!htb]
\centering
\includegraphics[height=12cm]{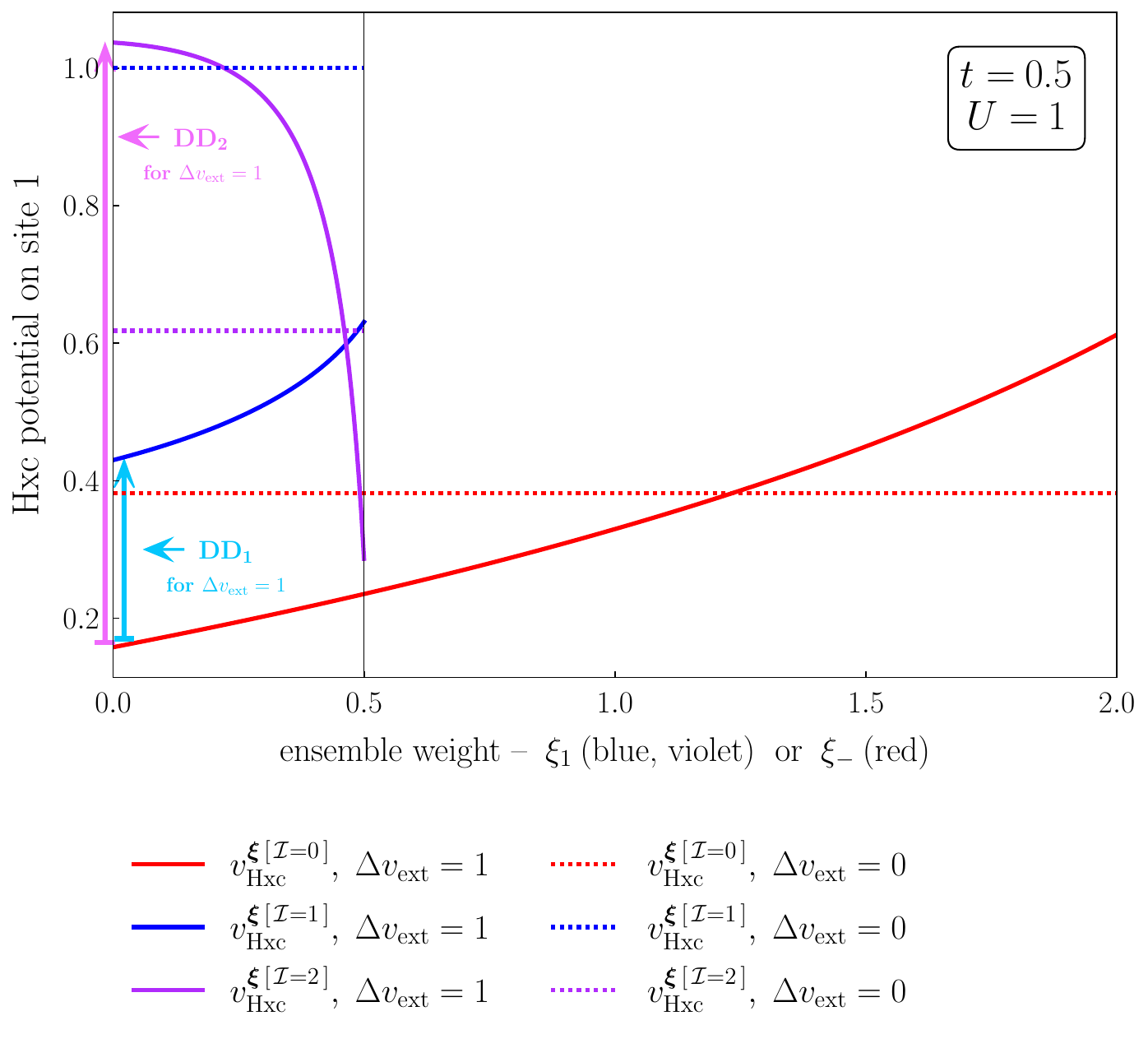} 
\caption{Exact Hxc potential on site $1$ (see Eqs.~(\ref{eq:Hxc_pot_eq_KS_minus_ext_HD}) and (\ref{eq:Hxc_pot_with_constant_unique_HD})) plotted as a function of an ensemble weight ($\xi_-$ or $\xi_1$, depending on the considered ionization process) for symmetric (dotted lines) and asymmetric ($\Delta v_{\rm ext}=1$, solid lines) Hubbard dimers with $U/t=2$. For the ionization of the ground state ($\mathcal{I}=0$, red curves), $\xi_-$ varies in the range $0< \xi_-\leq 2$ while $\xi_1=\xi_2=0$ (see Eqs.~(\ref{eq:range_xi2}) and (\ref{eq:range_xi1})). As for the ionization of the first excited state ($\mathcal{I}=1$, blue curves), $\xi_-\rightarrow 0^+$, $\xi_2=0$, and $\xi_1$ varies in the range $0<\xi_1< 1/2$. The Hxc potential associated with the ionization of the second excited state ($\mathcal{I}=2$, violet curves) is also plotted as a function of $\xi_1$ (in the range $0<\xi_1< 1/2$) for $\xi_-\rightarrow 0^+$ and $\xi_2\rightarrow 0^+$. The vertical cyan blue and magenta arrows show the derivative discontinuities that the Hxc potential exhibits when crossing regular two-electron ground-state DFT (where all the weights equal zero) from the ionized ground state toward the first and second excited states, respectively. 
}     
\label{fig:Fig1}   
\end{figure*} 


In a recent work~\cite{cernatic2024neutral}, we investigated the derivative discontinuity that the Hxc potential exhibits when the first singlet excited state is incorporated into the ensemble under study. For that purpose, we compared two scenarios which are reproduced in Fig.~\ref{fig:Fig1} in the moderately correlated $U/t=2$ regime, for both symmetric ($\Delta v_{\rm ext}=0$) and asymmetric ($\Delta v_{\rm ext}=1$) dimers. In the first scenario, where $0<\xi_-\leq 2$ and $\xi_1=\xi_2=0$, the Hxc potential is uniquely defined from the ionization of the two-electron ground state, previously labelled as $\mathcal{I}=0$ (see Eqs.~(\ref{eq:general_constraint_Koopmans_Hcv_pot}) and (\ref{eq:Dfunc_ionization_GS})). In the second scenario, where $\xi_-\rightarrow 0^+$, $0<\xi_1<1/2$, and $\xi_2=0$, the Hxc potential is defined from the ionization of the first excited state ($\mathcal{I}=1$), according to Eqs.~(\ref{eq:general_constraint_Koopmans_Hcv_pot}) and (\ref{eq:Dfunc_ionization_1stexcitation}). In this work, we focus on the modification of the Hxc potential when the ensemble density-functional KS system undergoes a double excitation (note that its connection with the excitation process that the true interacting system undergoes will be discussed in detail in the next section). For that purpose, we introduce a third scenario that differs from the second one only by the infinitesimal incorporation of the second excited state into the ensemble, \ie, $\xi_-\rightarrow 0^+$, $0<\xi_1<1/2$, and $\xi_2\rightarrow 0^+$, so that the Hxc potential can now be uniquely defined from the ionization of the latter state ($\mathcal{I}=2$), according to Eq.~(\ref{eq:Dfunc_ionization_2ndexcitation}). As shown in Fig.~\ref{fig:Fig1}, the Hxc potential does exhibit a derivative discontinuity when switching from the first to the second excited state, as expected from Eq.~(\ref{eq:DD_general_case}). We also note that, in the asymmetric case ($\Delta v_{\rm ext}=1$), the two Hxc potentials differ substantially in their variation with respect to the ensemble weight $\xi_1$, especially when approaching the $\xi_1=1/2$ limit. This can be rationalized as follows. According to the final Hxc potential expression (on site 1) given in Eq.~(\ref{eq:Hxc_pot_with_constant_unique_HD}), and Eq.~(\ref{eq:general_KS_pot_diff_HD}), the deviation in Hxc potential between the first and second excited states can be expressed exactly as follows,             
\be
\begin{split}
&v^{(0^+,\xi_1,0^+)[{2}]}_{{\rm Hxc}}-v^{(0^+,\xi_1,0)[{1}]}_{{\rm Hxc}}
\\
&\underset{0<\xi_1\leq 1/2}{=}
\left[\mathscr{D}^{\bxi[2]}_{\rm Hxc}(n^{\bxi})-\mathscr{D}^{\bxi[1]}_{\rm Hxc}(n^{\bxi})\right]_{\bxi=(0,\xi_1,0)},
\end{split}
\ee
or, equivalently (see Eqs.~(\ref{eq:Dfunc_ionization_1stexcitation}), and (\ref{eq:Dfunc_ionization_2ndexcitation})), 
\be\label{eq:diff_Hxc_pot2_pot1_HD}
\begin{split}
&v^{(0^+,\xi_1,0^+)[{2}]}_{{\rm Hxc}}-v^{(0^+,\xi_1,0)[{1}]}_{{\rm Hxc}}
\\
&=\left[\left.\dxinum{{E_{\rm Hxc}^{\bxi}(n)}}{2}\right|_{n=n^{\bxi}}-\left.\dxinum{{E_{\rm Hxc}^{\bxi}(n)}}{1}\right|_{n=n^{\bxi}}\right]_{\bxi=(0,\xi_1,0)},
\end{split}
\ee
where, according to the reduction in ensemble size discussed in appendix~\ref{app1} (see also Eqs.~(\ref{eq:effective_weights_HD}), (\ref{eq:effective_dens_HD}), and (\ref{eq:Hxc_func_from_triens_HD})),  
\be
\begin{split}
&\left.\dxinum{{E_{\rm Hxc}^{\bxi}(n)}}{2}\right|_{\bxi=(0,\xi_1,0)}=2U
\\
& 
-\Bigg[3E^{\bxi}_{\rm Hxc}(n)
-\dfrac{\partial \zeta_1({\bxi})}{\partial \xi_2}\dxinum{{E_{\rm Hxc}^{\bxi}(n)}}{1}
\\
&\hspace{1cm}-\dfrac{\partial\nu(n,{\bxi})}{\partial\xi_2}\dfrac{\partial E_{\rm
Hxc}^{\bxi}(n)}{\partial n}\Bigg]_{\bxi=(0,\xi_1,0)}
,
\end{split}
\ee
thus leading to (see Eq.~(\ref{eq:Hxc_pot_-minus_deriv_n_HD}))
\be\label{eq:diff_weight_deriv2_weight_deriv1_HD}
\begin{split}
&\left[\dxinum{{E_{\rm Hxc}^{\bxi}(n)}}{2}-\dxinum{{E_{\rm Hxc}^{\bxi}(n)}}{1}\right]_{\bxi=(0,\xi_1,0)}=2U
\\
& 
-\Bigg[3E^{\bxi}_{\rm Hxc}(n)
+(2-3\xi_1)\dxinum{{E_{\rm Hxc}^{\bxi}(n)}}{1}
\\
&\hspace{1cm}+3(n-1)\Delta v_{\rm Hxc}^{\bxi}(n)\Bigg]_{\bxi=(0,\xi_1,0)}
.
\end{split}
\ee
As readily seen from Eqs.~(\ref{eq:diff_Hxc_pot2_pot1_HD}) and (\ref{eq:diff_weight_deriv2_weight_deriv1_HD}), the difference in Hxc potentials consists of three density-functional contributions to which $2U$ is added. One of them, which reads more explicitly as follows,
\begin{subequations}
\begin{align}
&-\left[\left.3(n-1)\Delta v_{\rm Hxc}^{\bxi}(n)\right|_{n=n^{\bxi}}\right]_{\bxi=(0,\xi_1,0)}
\\
\nonumber
&=-\left[\left.3(n-1)\Delta v_{\rm KS}^{\bxi}(n)\right|_{n=n^{\bxi}}\right]_{\bxi=(0,\xi_1,0)}
\\
\label{eq:KS_pot_contr_diff_pots}
&\quad+3(n^{\xi_1}-1)\Delta v_{\rm ext}
\\
\label{eq:explicit_expression_KS_pot_contr_diff_pots}
&=-\dfrac{6t(n^{\xi_1}-1)^2}{\sqrt{(1-\xi_1)^2 - (1-n^{\xi_1})^2}}+3(n^{\xi_1}-1)\Delta v_{\rm ext},
\end{align}
\end{subequations}
where we used the shorthand notation $n^{\xi_1}:=n^{\bxi=(0,\xi_1,0)}$,
relates to the ensemble KS potential (see Eqs.~(\ref{eq:general_KS_pot_diff_HD}) and (\ref{eq:KS_pot_contr_diff_pots})). In the asymmetric $U=2t=\Delta v_{\rm ext}=1$ regime depicted in Fig.~\ref{fig:Fig1}, the ensemble density $n^{\xi_1}$ varies weakly with $\xi_1$ in the range $1.4\leq n^{\xi_1} <1.5$~\cite{deur2017exact,cernatic2024neutral}. This explains why the Hxc potential for the second excited state ($\mathcal{I}=2$) decreases sharply with $\xi_1$ when approaching the limit $\xi_1=1/2$ (see the denominator in the first term of Eq.~(\ref{eq:explicit_expression_KS_pot_contr_diff_pots})).\\ 

Let us finally note that, when the dimer is symmetric (\ie, $\Delta v_{\rm ext}=0$), the ensemble density equals $n^{\bxi}=1$ and~\cite{cernatic2024neutral} 
\be
\begin{split}
\left.E^{\bxi}_{\rm Hxc}(n)\right|_{n=1}&\overset{\bxi=(0,\xi_1,0)}{=}\dfrac{U(1+\xi_1)}{2}
\\
&\hspace{1cm}+(1-\xi_1)\left(2t-\frac{1}{2}\sqrt{U^2+16t^2}\right),
\end{split}
\ee
so that (see Eq.~(\ref{eq:Hxc_func_from_triens_HD}))  
\be
\begin{split}
\left.E^{\bxi}_{\rm Hxc}(n)\right|_{n=1}
&
\overset{\bxi=(0,\xi_1,\xi_2)}{=}\dfrac{U\left(1+\xi_1\right)}{2}
\\
&+\left(1-\xi_1-2\xi_2\right)\left(2t-\frac{1}{2}\sqrt{U^2+16t^2}\right).
\end{split}
\ee
Thus we conclude that, like the Hxc potential defined from the ionization of the first excited state~\cite{cernatic2024neutral}, the one deduced from the ionization of the second excited state is weight-independent and it deviates from the latter as follows, according to Eq.~(\ref{eq:diff_Hxc_pot2_pot1_HD}),
\be
v^{(0^+,\xi_1,0^+)[{2}]}_{{\rm Hxc}}-v^{(0^+,\xi_1,0)[{1}]}_{{\rm Hxc}}\underset{\Delta v_{\rm ext}=0}{=}-2t-\dfrac{U}{2}+\frac{1}{2}\sqrt{U^2+16t^2},
\ee
which is in perfect agreement with Fig.~\ref{fig:Fig1}.

\subsubsection{Analysis of the physical eigenstates in the ensemble density-functional KS representation}\label{sec:analysis_WFs_KS_rep_HF}

The infinitesimal $\xi_-\rightarrow 0^+$ incorporation of the ionized ground state, which was essential for describing derivative discontinuities in the previous section, is of no use in the following discussion since we are interested in the physical and KS states, which are invariant under any uniform shift in potential. Therefore, we can simply set $\xi_-=0$ and study the regular TGOK ensemble consisting of the two-electron ground state and the two lowest (singlet) neutral excited states (with weights $\xi_1$ and $\xi_2$, respectively). The (weight-independent) physical eigenstates can be decomposed as follows in the lattice (\ie, localized) representation,
\be\label{eq:decom_wfs_lattice_CSFs_HD}
\myket{\Psi_\nu} = \sum^2_{K=0}\langle \Xi_K\vert\Psi_\nu\rangle \myket{\Xi_K}, \hspace{0.2cm}\nu=0,1,2,
\ee
where~\cite{senjean2017local}
\begin{subequations}\label{eq:lattice_CSFs_basis_HD}
\begin{align}
\myket{\Xi_0}&=\cdop_{0\uparrow}\cdop_{0\downarrow}\myket{\rmvac},
\\
\myket{\Xi_1}&=\cdop_{1\uparrow}\cdop_{1\downarrow}\myket{\rmvac},
\\
\myket{\Xi_2}&=\dfrac{1}{\sqrt{2}}\left(\cdop_{0\uparrow}\cdop_{1\downarrow}-\cdop_{0\downarrow}\cdop_{1\uparrow}\right)\myket{\rmvac}.
\end{align}
\end{subequations}
We are in fact interested in the representation of the eigenstates in the ({\it a priori} weight-dependent, according to Eq.~(\ref{eq:ens_KS_eqs_many-body_KS_states})) ensemble density-functional KS basis, \ie,
\be
\myket{\Psi_\nu}=\sum^2_{\mu=0}\langle \Phi^{\bxi}_\mu\vert\Psi_\nu\rangle\myket{\Phi^{\bxi}_\mu}, \hspace{0.2cm}\nu=0,1,2.
\ee
The derivation of both representations is discussed in detail in Appendix~\ref{app:CI_coefficients}.

\begin{figure*}[!t]
\centering
\includegraphics[height=16cm]{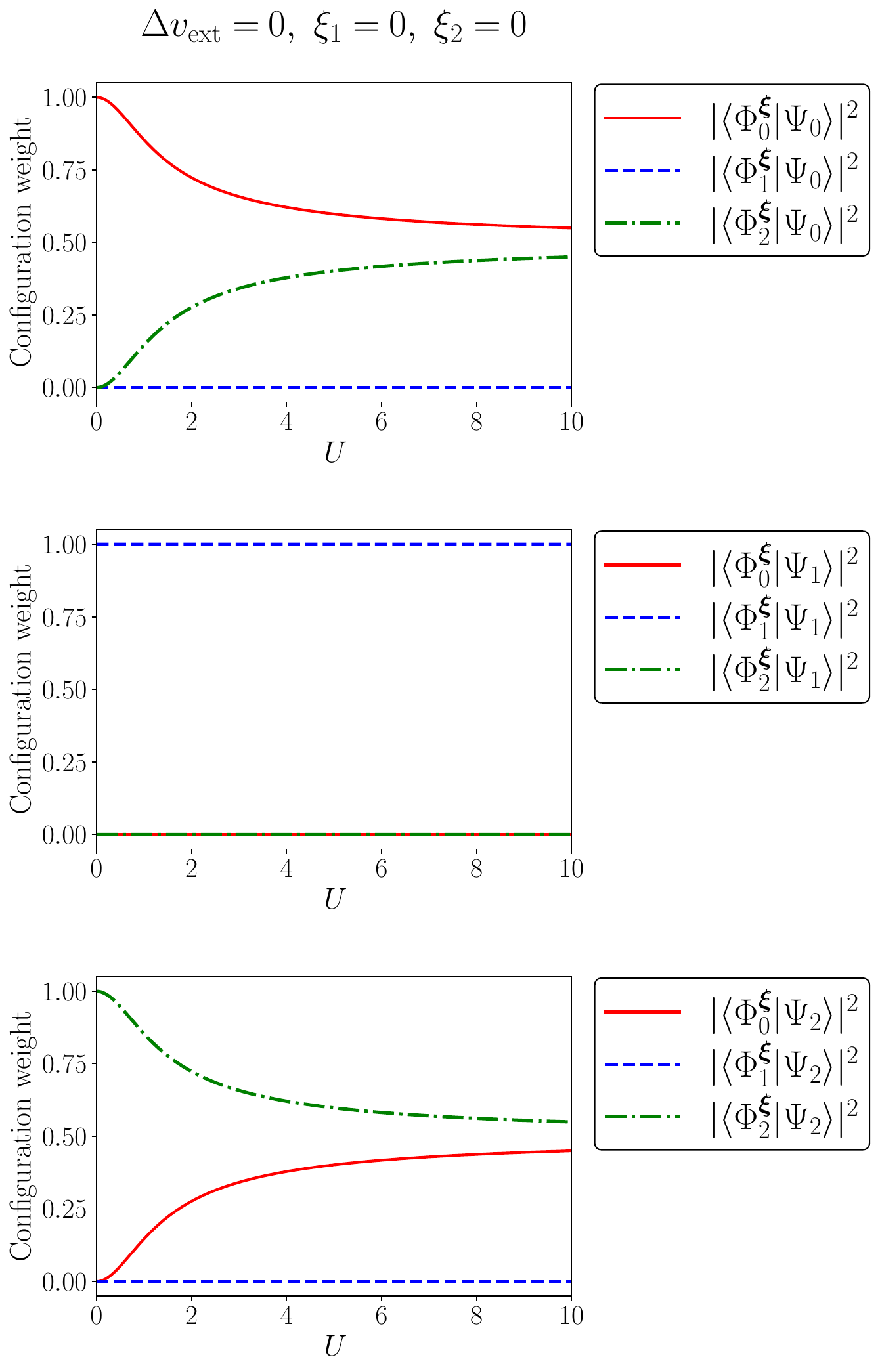} 
\caption{Configuration
weights in the ensemble density-functional KS basis $\left\{\Phi^{\bxi}_\mu\right\}$ of the true interacting eigenfunctions $\left\{\Psi_\nu\right\}$ plotted as functions of the interaction
strength $U$ in the symmetric Hubbard dimer ($\Delta v_{\rm ext} = 0$). The KS basis is weight-independent in this case. Top panel: ground-state expansion ($\nu=0$). Middle panel:  first excited state ($\nu=1$). Bottom panel: second excited state ($\nu=2$). See text for further details.}     
\label{fig:Fig2}   
\end{figure*} 

\begin{figure*}[!t]
\centering
\includegraphics[height=16cm]{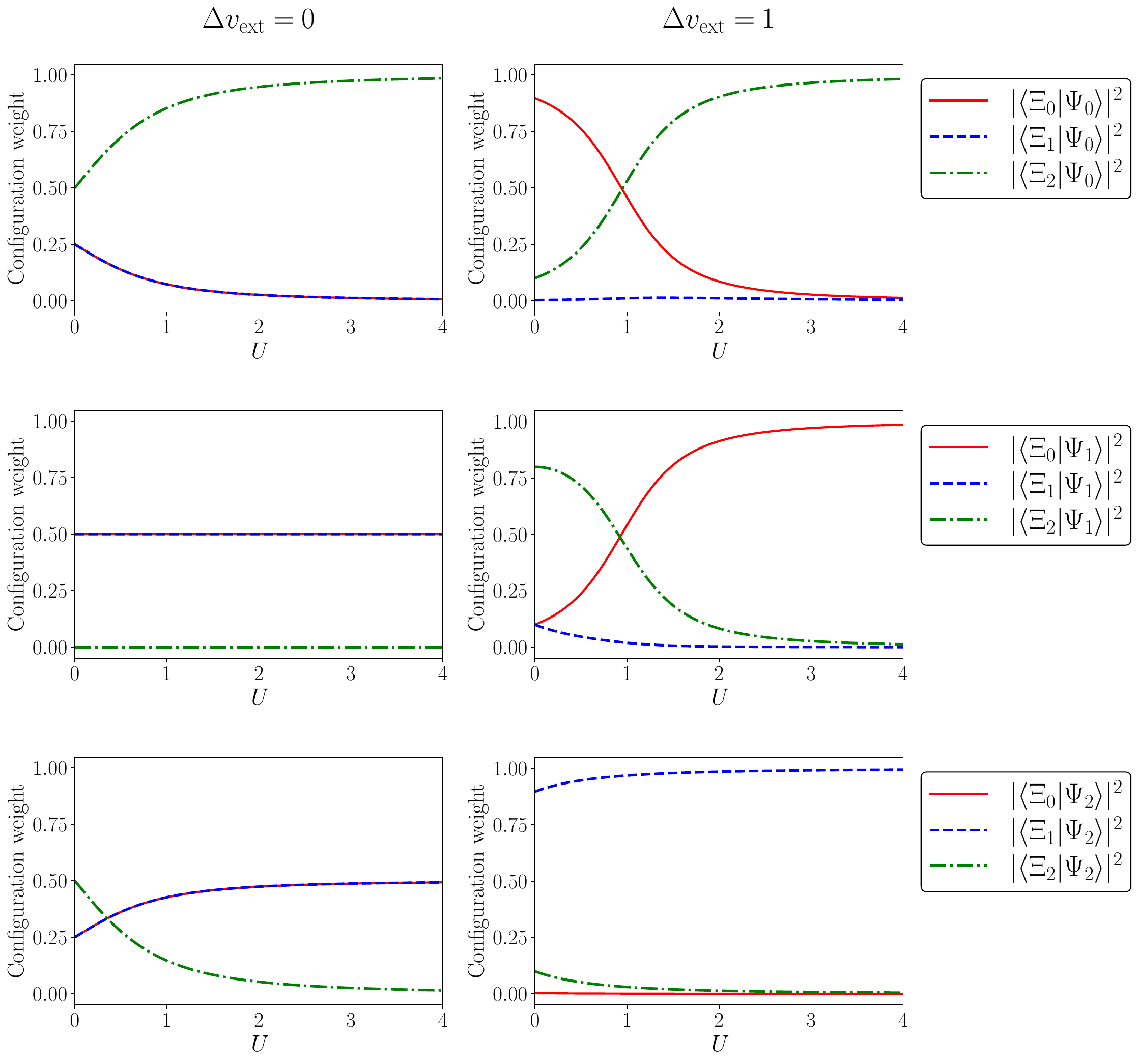} 
\caption{
Configuration
weights in the (local) site-based representation $\left\{\Xi_K\right\}$ of the true interacting eigenfunctions $\left\{\Psi_\nu\right\}$ plotted as functions of the interaction
strength $U$ in the symmetric ($\Delta v_{\rm ext} = 0$, left panels) and asymmetric ($\Delta v_{\rm ext} = 1$, right panels)  Hubbard dimers. Top panels: ground-state expansion ($\nu=0$). Middle panels:  first excited state ($\nu=1$). Bottom panels: second excited state ($\nu=2$). See text for further details.
}     
\label{fig:Fig3}   
\end{figure*}

Let us first consider the symmetric dimer ($\Delta v_{\rm ext}=0$). Since, in this case, the ensemble density equals 1~\cite{deur2017exact}, the KS potential difference equals zero. Consequently, the KS states are weight-independent and equivalent to the solutions of the regular H\"{u}ckel (or tight binding) problem for the hydrogen molecule in a minimal basis. The configuration weights of the interacting eigenstates in the KS representation are plotted in Fig.~\ref{fig:Fig2} as functions of $U$ (we recall that $t=1/2$ throughout this work). For analysis purposes, the configuration weights obtained in the lattice representation (\ie, in the basis of the $1s$ atomic orbitals if we pursue the analogy with the hydrogen molecule) are also plotted in the left panels of Fig.~\ref{fig:Fig3}. For symmetry reasons, the first (singlet) excited state is $U$-independent (it equals $\frac{1}{\sqrt{2}}(\myket{\Xi_0}-\myket{\Xi_1})$ and its energy is $U$) and, therefore, it matches the singly-excited KS state. On the other hand, as $U$ increases, both ground and second excited states (which belong to the same spatial symmetry) become mixtures of ground and doubly-excited KS states, as expected. Referring to the second excited state as ``doubly-excited'' is relevant in this case but we should remember that the ground-state KS configuration contributes significantly and, ultimately, equally, when the symmetric dimer becomes strictly correlated (\ie, when the hydrogen molecule dissociates).      

\begin{figure*}[!t]
\centering
\includegraphics[height=16cm]{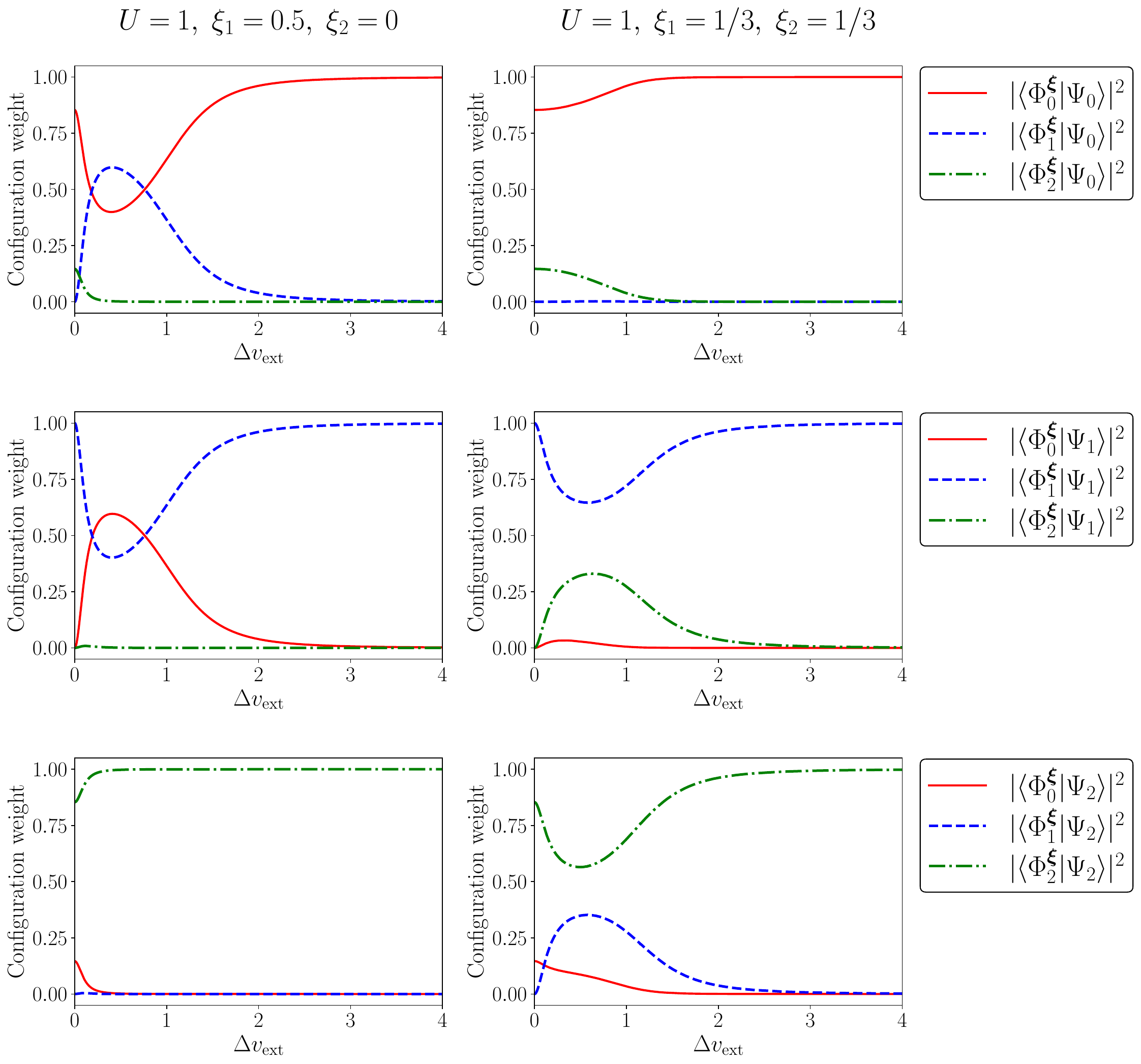} 
\caption{
Same as Fig.~\ref{fig:Fig2} but the configuration weights are now plotted for $U=1$ (and $t=1/2$) as functions of $\Delta v_{\rm ext}$ in the equi-bi- (left panels) and equi-tri-ensemble (right panels) density-functional KS representations ($\xi_-=0$ in both cases). See text for further details.
}     
\label{fig:Fig4}   
\end{figure*}

\begin{figure*}[!t]
\centering
\includegraphics[height=16cm]{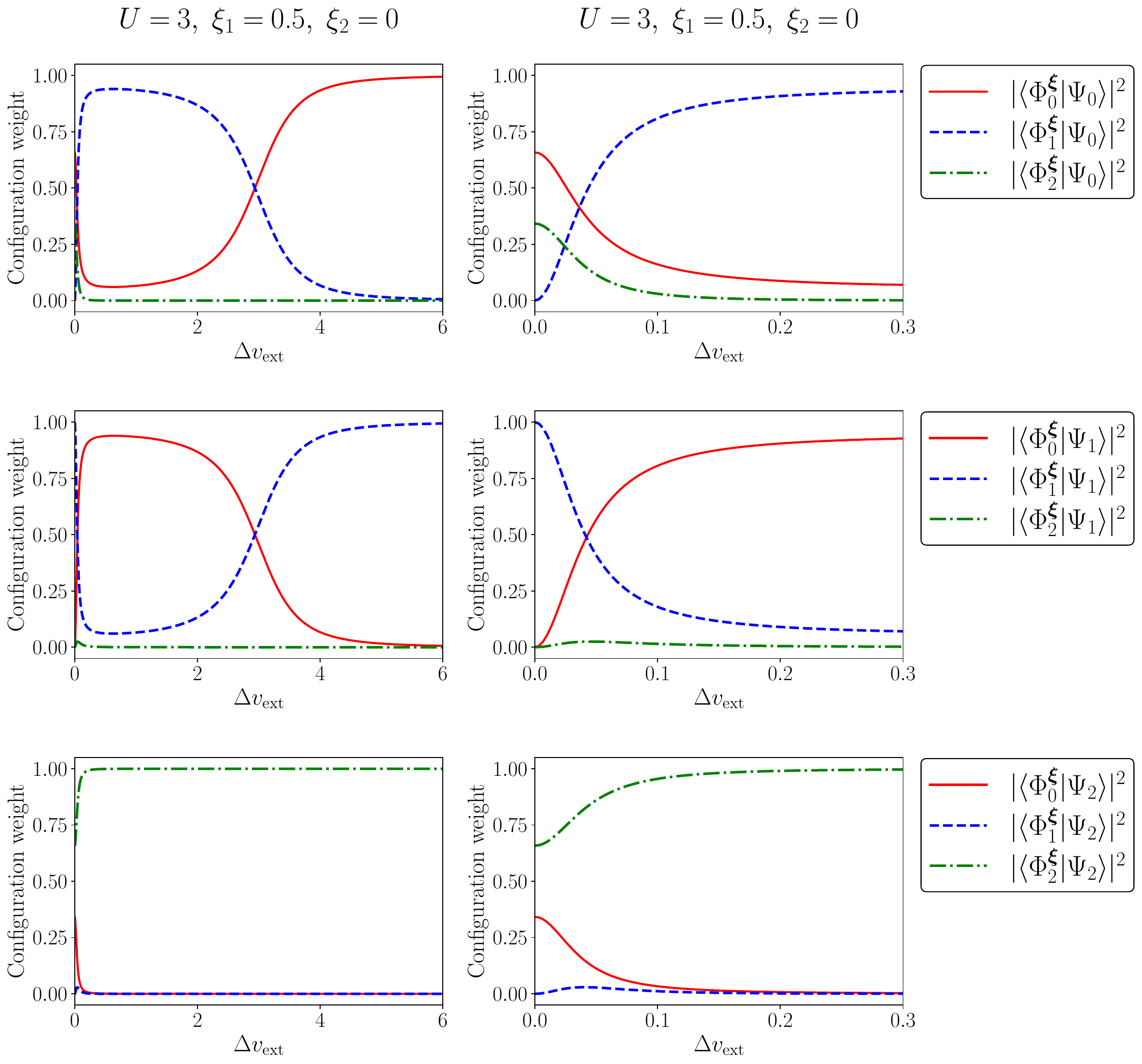} 
\caption{
Same as Fig.~\ref{fig:Fig4} for $U=3$ (and $t=1/2$) in the equi-bi-ensemble density-functional KS representation only. Right panels show details of the left panels in the range $0\leq \Delta v_{\rm ext}\leq 0.3$.}     
\label{fig:Fig5}   
\end{figure*}

\begin{figure*}[!t]
\centering
\includegraphics[height=16cm]{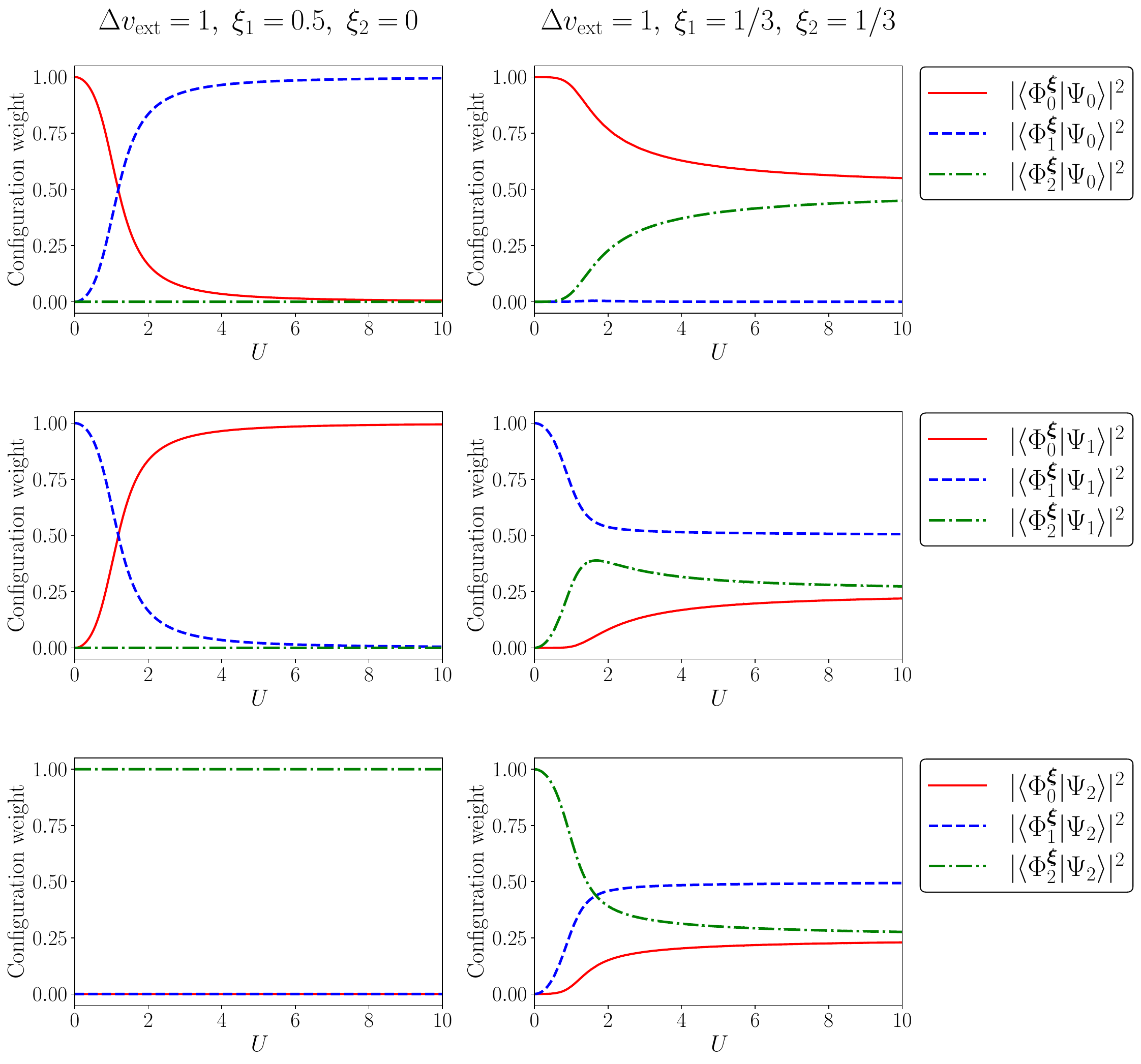} 
\caption{
Same as Fig.~\ref{fig:Fig2} for the asymmetric ($\Delta v_{\rm ext} = 1$) dimer in the 
equi-bi- (left panels) and equi-tri-ensemble (right panels) density-functional KS representations ($\xi_-=0$ in both cases). See text for further details.}     
\label{fig:Fig6}   
\end{figure*}

\begin{figure*}[!t]
\centering
\includegraphics[height=16cm]{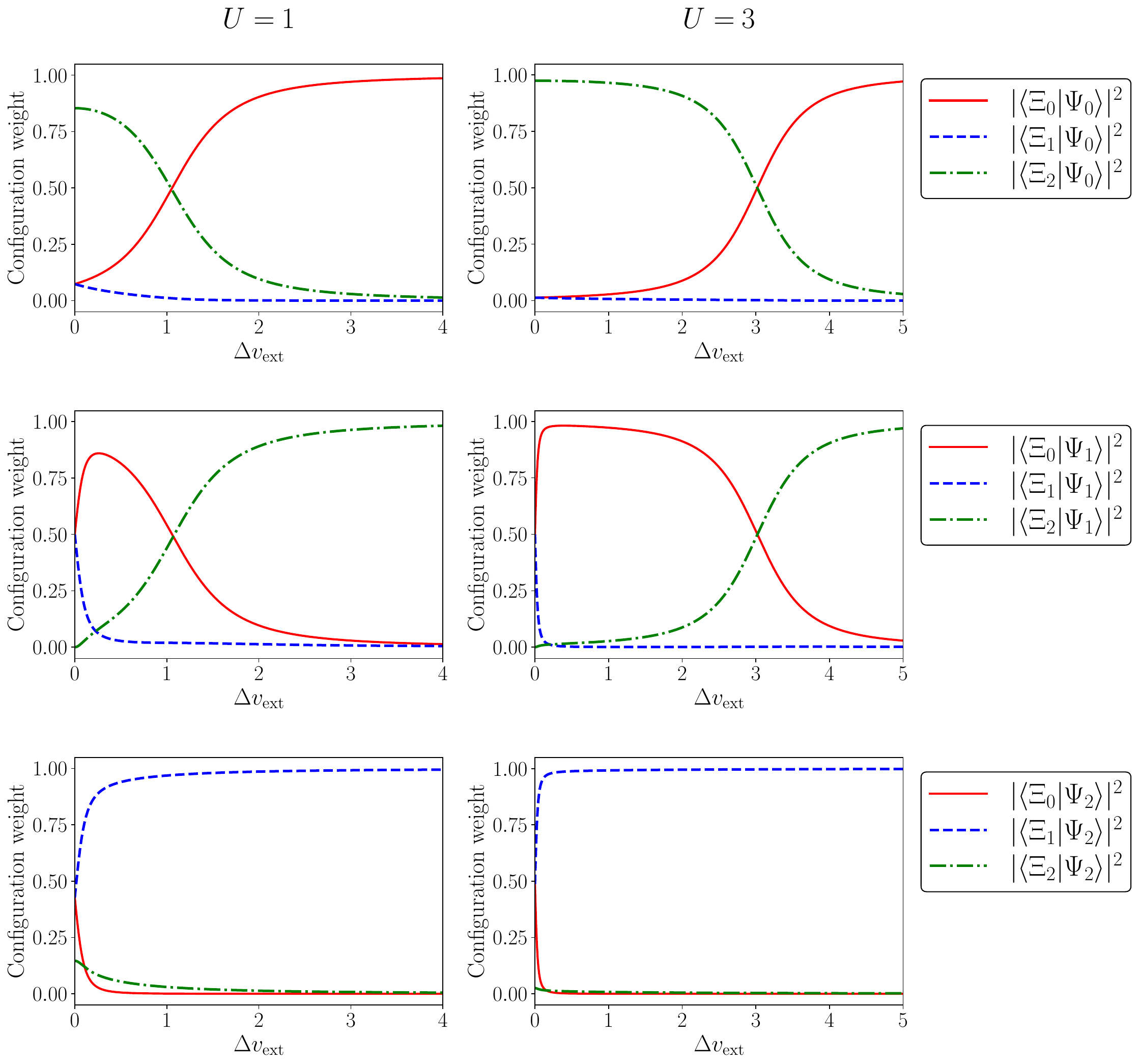} 
\caption{Same as Fig.~\ref{fig:Fig3} but the (localized) site-based configuration weights are now plotted for $U=1$ (left panels) and $U=3$ (right panels) as functions of $\Delta v_{\rm ext}$ with $t=1/2$.}     
\label{fig:Fig7}   
\end{figure*}

The impact of asymmetry on the interacting ground- and excited-state configuration expansions within the (now weight-dependent) ensemble density-functional KS representation is investigated in the moderately correlated $U/t=2$ regime in Fig.~\ref{fig:Fig4}. The stronger $U/t=6$ correlation regime is investigated in Fig.~\ref{fig:Fig5}. We focus here on equi-ensembles~\cite{ding2024ground}, which are commonly used in wavefunction theory calculations. Let us first consider the bi-ensemble density-functional case, {\ie}, $\xi_2=0$ and $\xi_1=\xi_0=1/2$ (see the left panels of both Figures). As soon as we slightly deviate from the symmetric case (\ie, for $\Delta v_{\rm ext}>0$), the second excited state (which does not belong to the bi-ensemble) rapidly reduces to the doubly-excited (bi-ensemble density-functional) KS determinant, as $\Delta v_{\rm ext}$ increases (see the bottom left panel of Fig.~\ref{fig:Fig4} and the bottom panels of Fig.~\ref{fig:Fig5}). On the other hand, for $U=1$, both ground and first excited states are mixtures of ground and singly-excited KS states, in the range $0<\Delta v_{\rm ext}\leq 3$. Referring to the first excited state as singly-excited is relevant in this case but we should of course remember that, because of electron correlation, the ground-state KS configuration may contribute significantly. Actually, in the vicinity of $\Delta v_{\rm ext}=t=1/2$, we notice that the latter contributes even more than the first excited KS one (see the top and middle left panels of Fig.~\ref{fig:Fig4}). In the stronger $U=3$ correlation regime, this feature is even more pronounced when $0.1<\Delta v_{\rm ext}<U$ (see the top and middle panels of Fig.~\ref{fig:Fig5}). For completeness, we plot in the left panels of Fig.~\ref{fig:Fig6} the configuration weights as functions of $U$ for the fixed $\Delta v_{\rm ext}=1$ asymmetric potential value. As readily seen from the top and middle panels, as we approach the strictly correlated $U/t\rightarrow +\infty$ limit, the physical interacting eigenstates become pure KS states with a major difference though: The first excited state turns out to be the ground KS state, and {\it vice versa}. The reason is the following. In this regime, the ground- and first excited-state densities are close to 1 (because the ground-state wavefunction is essentially that of the strongly correlated and symmetric dimer) and 2, respectively, as deduced from the top and middle right panels of Fig.~\ref{fig:Fig3} (see also Eqs.~(\ref{eq:decom_wfs_lattice_CSFs_HD}) and (\ref{eq:lattice_CSFs_basis_HD})). Consequently, the equi-bi-ensemble density is close to 1.5, which means that the equi-bi-ensemble KS potential difference is approaching $+\infty$ (see Eq.~(\ref{eq:general_KS_pot_diff_HD})). As a result, in the ground KS state, the two electrons are essentially localized on site 0, which corresponds to the {\it first} interacting excited state. On the other hand, in the first excited KS state, the density equals 1 on both sites, exactly like in the interacting {\it ground} state. We note finally that, in the strongly asymmetric $\Delta v_{\rm ext}>> U$ regimes depicted in the left panels of Figs.~\ref{fig:Fig4} and \ref{fig:Fig5}, physical and KS states become essentially identical as $\Delta v_{\rm ext}$ approaches $+\infty$. Indeed, in this regime, the equi-bi-ensemble density is still close to 1.5, as deduced from the top and middle panels of Fig.~\ref{fig:Fig7}. Therefore, the KS states are unchanged but the interacting ground state now consists of two electrons localized on site 0 while the first excited state has a density equal to 1 on both sites, exactly like in the KS world.

Let us now turn to the equi-tri-ensemble density-functional case. As clearly illustrated in Figs.~\ref{fig:Fig4} and \ref{fig:Fig5}, moving from a bi- to a tri-ensemble completely changes the ensemble density-functional KS basis and, therefore, the representation of the physical eigenstates (which are unchanged) in the latter basis. For the fixed $U=1$ interaction strength value, the doubly-excited KS state contributes to both ground and first excited interacting states for a broader range of $\Delta v_{\rm ext}$ values. In the latter asymmetric regime, we also notice that, unlike in the equi-bi-ensemble case, the ground KS state gives a relatively good description of the true ground state (see the top panels of Fig.~\ref{fig:Fig4}), while both first and second excited states are mixtures of singly- and doubly-excited KS states (see the middle and bottom right panels of Fig.~\ref{fig:Fig4}). The overall change in ensemble density-functional KS representation of the true eigenstates, when moving from a bi- to a tri-equi-ensemble, can be rationalized as follows. As pointed out in Sec.~\ref{sec:DD_doubles_HD}, when $U=2t=\Delta v_{\rm ext}=1$, for example, the equi-bi-ensemble density is relatively close to 1.5 (both ground- and first excited-state densities are close to the latter value~\cite{deur2017exact}), which means that the equi-bi-ensemble KS potential (for which $\xi_1=1/2$ and $\xi_2=0$ in Eq.~(\ref{eq:general_KS_pot_diff_HD})) is very attractive on site 0. Therefore, in this case, the KS ground state essentially consists of two electrons localized on site 0, which does not reflect at all the true ground-state electronic structure (see the top left panel of Fig.~\ref{fig:Fig7}). On the other hand, when $\xi_1=1/3$ and $\xi_2=(1/3)-\eta$, where $\eta\rightarrow 0^+$, so that we can approach the equi-tri-ensemble case, the density equals 
\begin{subequations}
\begin{align}
\label{eq:tri-ens_dens_eta_HD}
n^{\rm tri}&=\left(\frac{1}{3}+\eta\right)n_{\Psi_0}+\frac{1}{3}n_{\Psi_1}+\left(\frac{1}{3}-\eta\right)n_{\Psi_2}
\\
&\approx 1+3\eta \left(n_{\Psi_0}-1\right), 
\end{align}
\end{subequations}
where we used the fact that $n_{\Psi_2}=3-n_{\Psi_0}-n_{\Psi_1}$~\cite{deur2019ground} and, in the considered regime, $n_{\Psi_0}\approx n_{\Psi_1}$.~\cite{deur2017exact} Consequently, the KS potential difference can be simplified as follows (see Eq.~(\ref{eq:general_KS_pot_diff_HD})),
\be
\Delta v^{\rm tri}_{\rm KS}\approx \dfrac{6t\eta \left(n_{\Psi_0}-1\right)}{\sqrt{4\eta^2-(n^{\rm tri}-1)^2}}=\dfrac{6t\left(n_{\Psi_0}-1\right)}{\sqrt{4-9\left(n_{\Psi_0}-1\right)^2}}.
\ee
As readily seen from the above equation, unlike in the equi-bi-ensemble case, the KS potential does not become singular when $n_{\Psi_0}$ is approaching 1.5, which is the case in the considered regime. Therefore, the KS potential is now much less attractive on site 0 and the electrons are more delocalized in the KS ground state, like in the interacting ground state. 


Note that, in this moderately correlated case, each KS excited state still gives a qualitatively correct description of each physical excited state.   
In the stronger $1=\Delta v_{\rm ext}<<U$ correlation regime, where the equi-bi-ensemble density is even closer to 1.5 (since $n_{\Psi_0}\approx 1$ and $n_{\Psi_1}\approx 2$~\cite{deur2017exact}, thus leading to $n_{\Psi_2}\approx 0$), the equi-tri-ensemble density reduces to $n^{\rm tri}\approx 1+\eta$ (see Eq.~(\ref{eq:tri-ens_dens_eta_HD})) and
\be
\Delta v^{\rm tri}_{\rm KS}\approx \dfrac{2t\eta}{\sqrt{4\eta^2-(n^{\rm tri}-1)^2}}=\dfrac{2t}{\sqrt{3}},
\ee
which is again finite, unlike the equi-bi-ensemble KS potential difference which tends to $+\infty$. This explains the drastic change in representation of the interacting eigenstates when moving from the bi- to the tri-ensemble case (see the left and right panels of Fig.~\ref{fig:Fig6}). For example, the true ground state is described, for large $U$ values, through an equal mixing of ground and doubly-excited KS states, like in the strongly correlated symmetric dimer. On the other hand, both first and second excited states are combinations of ground (25\%), singly-excited (50\%), and doubly-excited (25\%) KS states. In this strongly correlated regime, the one-particle picture of electronic excitations completely breaks down, as expected, thus making labels such as "single excitation" or "double excitation" irrelevant for the true physical excitation processes.

\section{Conclusions and outlook}\label{sec5}

As a complement to a recent work~\cite{cernatic2024neutral}, where the e$N$c ensemble density functional theory of electronic excitations has been introduced, we extended in the present paper the theory to the description of single-electron excitations from {\it any} occupied orbital in the KS ground state and, most importantly, to the challenging double excitations. The exactification of Koopmans' theorem for {\it single-electron} ionization processes and the related concept of density-functional Hxc derivative discontinuity, whose mathematical construction fully relies on the {\it weight-dependent} ensemble density-functional Hxc energy, still play a central role. The theory has been implemented within the two-electron Hubbard dimer model. Nontrivial modifications of the exact Hxc potential upon neutral excitation processes, including the expected derivative discontinuities, have been highlighted and rationalized. Finally, in order to clarify the statement ``ensemble DFT can describe double excitations'', and also discuss what labels like ``single excitation" or ``double excitation" actually mean in the context of ensemble DFT, we have analyzed the representation of the three lowest two-electron (singlet) eigenstates of the Hubbard dimer in both equi-bi- and equi-tri-ensemble density-functional KS bases. Even though the true interacting and KS ensembles share the same density, they can be drastically different. In some regimes, the states can be similar but their ordering in energy is different. In some other regimes, the physical states are mixtures of ground and excited KS states. This analysis also reveals that the KS representation of the physical eigenstates can be very sensitive to the choice of ensemble, through its dependence on the ensemble density. While the present work focused on the exact theory, the next challenging task consists in developing density-functional approximations in this context. Combining a generalized KS formulation of the theory with perturbative ensemble DFT~\cite{gould2022single} would, for example, be an interesting path to follow. We may also learn from the time-dependent linear response of density-functional ensembles. Indeed, like the static formulation of ensemble DFT, the latter response is expected to give us access to the excitation energies. Work is currently in progress in these different directions.

\section*{Acknowledgements}
The authors thank ANR (CoLab project, grant no.: ANR-19-CE07-0024-02)
for funding as well as P.-F. Loos and B. Senjean for fruitful discussions.

\appendix

\section{Computation of exact ensemble density functionals and reduction to a tri-ensemble}\label{app1}
\vspace*{12pt}

Throughout Sec.~\ref{sec:Application},
we employ the density
functional $F^{\bxi}(n)$ introduced in Eq.~(\ref{eq:LL_def_ens_Hxc_func}) for an 
e$N$c ensemble consisting of 
the ground, first and second excited (singlet) two-electron
states, and the cationic ground one-electron
state. It is characterized by the
collection of weights
${\bxi = (\xi_-,\xi_1,\xi_2)}$. The functional $F^{\bxi}(n)$ is the analog for e$N$c ensembles of the Levy--Lieb functional~\cite{levy1979universal,LFTransform-Lieb}, that we assume to be equivalent to a Lieb functional~\cite{LFTransform-Lieb} for densities under study, like in Ref.~\cite{cernatic2024neutral}. Consequently, it can be evaluated through a Legendre--Fenchel transform, as follows,
\be\label{eq:4-state_Lieb_func_HD}
\begin{aligned}
F^{\bxi}(n) = \sup_{\Delta v}\biggl\{
&\left(1-\dfrac{\xi_-}{2}-\xi_1-\xi_2\right)E_0^N(\Delta v)
\\
&
+ \xi_- E_0^{N-1}(\Delta v) 
+ \xi_1 E_1^N(\Delta v) 
\\
&+ \xi_2 E_2^N(\Delta v)
+{\Delta v}(n-1)
\biggr\} .  
\end{aligned}
\ee
Since the (singlet) two-electron energies in the Hubbard dimer
sum up to $2U$~\cite{deur2019ground}, we can afford the reduction of the above four-state
ensemble to an effective tri-ensemble (consisting of the ground, first excited, and cationic states) by substituting 
${E_2^N(\Delta v) = 2U - E_1^N(\Delta v) - E_0^N(\Delta v)}$ into
the above equation. This reduction is completely analogous to
the TGOK tri-to-bi-ensemble reduction in Ref.~\citenum{deur2019ground} (see Eqs.~(A2)-(A4) therein), leading to a similar expression for $F^{\bxi}(n)$,
which reads as follows,
\be
F^{\bxi}(n) = 2U\xi_2 + (1-3\xi_2)F^{\bzeta}(\nu),
\ee
where $\bzeta = (\zeta_-,\zeta_1)$ is the 
collection of weights for the tri-ensemble, with the two effective weights equal to ${\zeta_- = \xi_-/(1-3\xi_2)}$ and
${\zeta_1 = (\xi_1 - \xi_2)/(1-3\xi_2)}$. The effective tri-ensemble density $\nu$ reads
${\nu = (n-3\xi_2)/(1-3\xi_2)}$. The tri-ensemble Lieb functional $F^{\bzeta}(\nu)$, which has been extensively used in Ref.~\citenum{cernatic2024neutral}, can be evaluated as follows,
\be
\begin{aligned}
F^{\bzeta}(\nu) = \sup_{\Delta v}\biggl\{
&\left(1-\dfrac{\zeta_-}{2}-\zeta_1\right)E_0^N(\Delta v)
+ \zeta_- E_0^{N-1}(\Delta v)
\\
&+\zeta_1 E_1^N(\Delta v)
+{\Delta v}(\nu-1)
\biggr\}.   
\end{aligned}
\ee

\section{Computation of wavefunction expansion coefficients in the lattice and ensemble KS representations}\label{app:CI_coefficients}

In the Hubbard dimer, the
singlet subspace of the two-electron Hilbert space comprises three configurations. In the
lattice site basis, they are expressed as follows,
\begin{subequations}
\begin{align}
\myket{\Xi_0}&=\cdop_{0\uparrow}\cdop_{0\downarrow}\myket{\rmvac},
\\
\myket{\Xi_1}&=\cdop_{1\uparrow}\cdop_{1\downarrow}\myket{\rmvac},
\\
\myket{\Xi_2}&=\dfrac{1}{\sqrt{2}}\left(\cdop_{0\uparrow}\cdop_{1\downarrow}-\cdop_{0\downarrow}\cdop_{1\uparrow}\right)\myket{\rmvac}.
\end{align}
\end{subequations}

Any singlet two-electron eigenstate can be expanded
in the basis of above configurations as follows,
\be\label{eq:Psi_I_site_expansion}
\myket{\Psi_\nu} = \sum^2_{K=0} C_{K\nu} \myket{\Xi_K}, \hspace{0.2cm}\nu=0,1,2,
\ee
where $\Psi_\nu\equiv \Psi_\nu^N$ ($N=2$ here), and $\left\{C_{K\nu}=\braket{\Xi_K|\Psi_\nu}\right\}$ are the expansion
coefficients, which can be obtained
by inserting Eq.~\eqref{eq:Psi_I_site_expansion} into the Schr\"{o}dinger equation
${\hat{H}\myket{\Psi_\nu}=E_\nu\myket{\Psi_\nu}}$ for the
ground, first and second excited states (${\nu=0,1,2}$, respectively), and projecting
into the site many-body basis
$\mybra{\Xi_{L}}$,
\be\label{eq:Psi_I_site_expansion_H_sitebasis}
\sum_{K}C_{K \nu}
\langle \Xi_L|\,\hat{H}\,| \Xi_K\rangle
= E_\nu C_{L \nu}.
\ee
Using Eq.~\eqref{eq:Hamiltonian_Hdim}
to evaluate the
Hamiltonian matrix elements
in the site basis (see also Ref.~\citenum{senjean2017local}), $\langle \Xi_L|\,\hat{H}\,| \Xi_K\rangle$,
\be
\begin{aligned}
\langle \Xi_0|\,\hat{H}\,| \Xi_0\rangle  &= U - \Delta v_{\rm ext},
\\
\langle \Xi_1|\,\hat{H}\,| \Xi_1\rangle  &= U + \Delta v_{\rm ext},
\\
\langle \Xi_2|\,\hat{H}\,| \Xi_2\rangle  &= 0,
\\
\langle \Xi_0|\,\hat{H}\,| \Xi_1\rangle  &= 0,
\\
\langle \Xi_0|\,\hat{H}\,| \Xi_2\rangle  &= \langle \Xi_1|\,\hat{H}\,| \Xi_2\rangle = -\sqrt{2}t,
\end{aligned}
\ee
we obtain from Eq.~\eqref{eq:Psi_I_site_expansion_H_sitebasis} a system of three linear equations for the coefficients $C_{K\nu}$,
\be\label{eq:C_matrix_coeffs}
\begin{aligned}
(U-\Delta v_{\rm ext} - E_\nu)C_{0\nu} - \sqrt{2}t C_{2\nu} = 0
\\
(U+\Delta v_{\rm ext} - E_\nu)C_{1\nu} - \sqrt{2}t C_{2\nu} = 0
\\
-\sqrt{2}t (C_{0\nu} + C_{1\nu}) - E_\nu C_{2\nu} = 0.
\end{aligned}
\ee
Assuming nondegeneracy,
we fix $C_{0\nu}$ and
express $C_{1\nu}$ and
$C_{2\nu}$ as follows,
\be
C_{1\nu} = \dfrac{U - \Delta v_{\rm ext} - E_\nu}{U + \Delta v_{\rm ext}  - E\nu}C_{0\nu}, \;\;\;\; C_{2\nu} = \dfrac{U-\Delta v_{\rm ext} - E_\nu}{\sqrt{2}t}C_{0\nu}.
\ee
Then, $C_{0\nu}$ is determined
by normalizing the squared norm of the coefficients to unity,
\be
\begin{aligned}
1&=\sum_{K=0}^2 |C_{K\nu}|^2
\\
&=|C_{0\nu}|^2
\left[
1+\left(\dfrac{U-\Delta v_{\rm ext} -E_\nu}{U+\Delta v_{\rm ext} - E_\nu}\right)^2
+ \dfrac{(U-\Delta v_{\rm ext} - E_\nu)^2}{2t^2}
\right]
\\
&=|C_{0\nu}|^2
\dfrac{
2t^2\left[(E_\nu - U - \Delta v_{\rm ext})^2 + (E_\nu - U + \Delta v_{\rm ext})^2\right]}{2t^2(U+\Delta v_{\rm ext} + E_\nu)^2}
\\
&+|C_{0\nu}|^2
\dfrac{
\left[(E_\nu - U)^2 - \Delta v_{\rm ext}^2\right]^2
}{2t^2(U+\Delta v_{\rm ext} + E_\nu)^2}.
\end{aligned}
\ee
Using the fact that
\be
(E_\nu - U - \Delta v_{\rm ext})^2 + (E_\nu - U + \Delta v_{\rm ext})^2 = 
2\left[(E_\nu-U)^2 + \Delta v_{\rm ext}^2 \right],
\ee
and, by introducing the following
notation,
\be
G_\nu = 
\left[(E_\nu - U)^2 - \Delta v_{\rm ext}^2\right]^2 + 
4t^2\left[(E_\nu-U)^2 + \Delta v_{\rm ext}^2 \right],
\ee
it follows that
\be
|C_{0\nu}|^2 = 
\dfrac{2t^2(U+\Delta v_{\rm ext} - E_\nu)}{G_\nu}.
\ee
For the coefficient $C_{0\nu}$, one may choose
the positive square root
of the above expression, which
gives
\be\label{eq:C0nu}
\begin{aligned}
C_{0\nu} &= 
\dfrac{\sqrt{2}t\,|U+\Delta v_{\rm ext} - E_\nu|}{\sqrt{G_\nu}}
\\
&=\sgn{(U+\Delta v_{\rm ext} - E_\nu)}
\dfrac{\sqrt{2}t\,(U+\Delta v_{\rm ext} - E_\nu)}{\sqrt{G_\nu}}.
\end{aligned}
\ee
Then, $C_{1\nu}$ and $C_{2\nu}$ can be expressed as follows,
\be\label{eq:C1nu}
\begin{aligned}
C_{1\nu} &= 
\dfrac{U-\Delta v_{\rm ext} - E_\nu}{U+\Delta v_{\rm ext} - E_\nu}
\dfrac{\sqrt{2}t\,|U+\Delta v_{\rm ext} - E_\nu|}{\sqrt{G_\nu}}
\\
&=\sgn{(U+\Delta v_{\rm ext} - E_\nu)}
\dfrac{\sqrt{2}t\,(U-\Delta v_{\rm ext} - E_\nu)}{\sqrt{G_\nu}},
\end{aligned}
\ee
\be\label{eq:C2nu}
\begin{aligned}
C_{2\nu} &= 
\dfrac{U-\Delta v_{\rm ext} - E_\nu}{\sqrt{2}t}
\dfrac{\sqrt{2}t\,|U+\Delta v_{\rm ext} - E_\nu|}{\sqrt{G_\nu}}
\\
&=\sgn{(U+\Delta v_{\rm ext} - E_\nu)}
\dfrac{(E_\nu-U)^2 - \Delta v_{\rm ext}^2}{\sqrt{G_\nu}}.
\end{aligned}
\ee
The above expressions for $C_{K\nu}$
are completely general.
For instance, in a
weight-dependent ensemble KS system (we consider the particular case where $\xi_-=0$, like in Sec.~\ref{sec:analysis_WFs_KS_rep_HF}), for which the
ground, singly- and doubly-excited 
KS wavefunctions (${\nu=0,1,2,}$ respectively) are expressed
in the site basis as follows,
\be\label{eq:Phi_I_site_expansion} 
\myket{\Phi_\nu^{\bxi}} = \sum^2_{K=0} D_{K\nu}^{\bxi} \myket{\Xi_K}, \hspace{0.2cm}\nu=0,1,2,
\ee
the expansion
coefficients $\left\{ D_{K\nu}^{\bxi}=\langle\Xi_K|\Phi_\nu^{\bxi}\rangle\right\}$
are obtained from
Eqs.~\eqref{eq:C0nu},
\eqref{eq:C1nu} and~\eqref{eq:C2nu}
by using the following substitutions,
\be
D_{K\nu}^{\bxi} \equiv C_{K\nu}\left(U=0,\Delta v_{\rm ext} \rightarrow \Delta v_{\rm KS}^{\bxi},
E_\nu \rightarrow \mathcal{E}_\nu^{\bxi} \right),
\ee
where $\Delta v_{\rm KS}^{\bxi}\equiv\Delta v_{\rm KS}^{\bxi}(n^{\bxi})$ is the ensemble density-functional KS potential difference (see Eq.~\eqref{eq:general_KS_pot_diff_HD}),
evaluated at the exact ensemble density $n^{\bxi}$, and $\left\{\mathcal{E}_\nu^{\bxi}\right\}$ are the
individual two-electron KS energies:
\begin{subequations}
\begin{align}
\mathcal{E}_0^{\bxi} &= -\sqrt{4t^2 + (\Delta v_{\rm KS}^{\bxi})^2} = -\mathcal{E}_2^{\bxi}, 
\\
\mathcal{E}_1^{\bxi} &= 0.
\end{align}
\end{subequations}
In Sec.~\ref{sec:analysis_WFs_KS_rep_HF}, we analyze the expansions
of interacting wavefunctions
in the ensemble KS basis:
\be
\myket{\Psi_\nu}=\sum^2_{\mu=0}\langle \Phi^{\bxi}_\mu\vert\Psi_\nu\rangle\myket{\Phi^{\bxi}_\mu}, \hspace{0.2cm}\nu=0,1,2.
\ee
The coefficients $\langle \Phi^{\bxi}_\mu\vert\Psi_\nu\rangle$ are simply obtained from Eq.~(\ref{eq:Psi_I_site_expansion}) as follows,
\be
\begin{aligned}
\langle \Phi^{\bxi}_\mu\vert\Psi_\nu\rangle
&= 
\sum_{K=0}^2
\langle \Phi^{\bxi}_\mu\vert
\Xi_K
\rangle
\langle
\Xi_K
\vert\Psi_\nu\rangle
\\
&=
\sum_{K=0}^2
D_{K\mu}^{\bxi}C_{K\nu}.
\end{aligned}
\ee




\begin{thebibliography}{76}%
\makeatletter
\providecommand \@ifxundefined [1]{%
 \@ifx{#1\undefined}
}%
\providecommand \@ifnum [1]{%
 \ifnum #1\expandafter \@firstoftwo
 \else \expandafter \@secondoftwo
 \fi
}%
\providecommand \@ifx [1]{%
 \ifx #1\expandafter \@firstoftwo
 \else \expandafter \@secondoftwo
 \fi
}%
\providecommand \natexlab [1]{#1}%
\providecommand \enquote  [1]{``#1''}%
\providecommand \bibnamefont  [1]{#1}%
\providecommand \bibfnamefont [1]{#1}%
\providecommand \citenamefont [1]{#1}%
\providecommand \href@noop [0]{\@secondoftwo}%
\providecommand \href [0]{\begingroup \@sanitize@url \@href}%
\providecommand \@href[1]{\@@startlink{#1}\@@href}%
\providecommand \@@href[1]{\endgroup#1\@@endlink}%
\providecommand \@sanitize@url [0]{\catcode `\\12\catcode `\$12\catcode
  `\&12\catcode `\#12\catcode `\^12\catcode `\_12\catcode `\%12\relax}%
\providecommand \@@startlink[1]{}%
\providecommand \@@endlink[0]{}%
\providecommand \url  [0]{\begingroup\@sanitize@url \@url }%
\providecommand \@url [1]{\endgroup\@href {#1}{\urlprefix }}%
\providecommand \urlprefix  [0]{URL }%
\providecommand \Eprint [0]{\href }%
\providecommand \doibase [0]{http://dx.doi.org/}%
\providecommand \selectlanguage [0]{\@gobble}%
\providecommand \bibinfo  [0]{\@secondoftwo}%
\providecommand \bibfield  [0]{\@secondoftwo}%
\providecommand \translation [1]{[#1]}%
\providecommand \BibitemOpen [0]{}%
\providecommand \bibitemStop [0]{}%
\providecommand \bibitemNoStop [0]{.\EOS\space}%
\providecommand \EOS [0]{\spacefactor3000\relax}%
\providecommand \BibitemShut  [1]{\csname bibitem#1\endcsname}%
\let\auto@bib@innerbib\@empty
\bibitem [{\citenamefont {Slater}(1930)}]{slater1930note}%
  \BibitemOpen
  \bibfield  {author} {\bibinfo {author} {\bibfnamefont {J.~C.}\ \bibnamefont
  {Slater}},\ }\href@noop {} {\bibfield  {journal} {\bibinfo  {journal} {Phys.
  Rev.}\ }\textbf {\bibinfo {volume} {35}},\ \bibinfo {pages} {210} (\bibinfo
  {year} {1930})}\BibitemShut {NoStop}%
\bibitem [{\citenamefont {Fock}(1930)}]{fock1930paperonhfmethod}%
  \BibitemOpen
  \bibfield  {author} {\bibinfo {author} {\bibfnamefont {V.}~\bibnamefont
  {Fock}},\ }\href@noop {} {\bibfield  {journal} {\bibinfo  {journal} {Z.
  Phys.}\ }\textbf {\bibinfo {volume} {61}},\ \bibinfo {pages} {126} (\bibinfo
  {year} {1930})}\BibitemShut {NoStop}%
\bibitem [{\citenamefont {Kohn}\ and\ \citenamefont
  {Sham}(1965)}]{kohn1965selfconsistent}%
  \BibitemOpen
  \bibfield  {author} {\bibinfo {author} {\bibfnamefont {W.}~\bibnamefont
  {Kohn}}\ and\ \bibinfo {author} {\bibfnamefont {L.~J.}\ \bibnamefont
  {Sham}},\ }\href@noop {} {\bibfield  {journal} {\bibinfo  {journal} {Phys.
  Rev.}\ }\textbf {\bibinfo {volume} {140}},\ \bibinfo {pages} {A1133}
  (\bibinfo {year} {1965})}\BibitemShut {NoStop}%
\bibitem [{\citenamefont {Loos}\ \emph {et~al.}(2019)\citenamefont {Loos},
  \citenamefont {Boggio-Pasqua}, \citenamefont {Scemama}, \citenamefont
  {Caffarel},\ and\ \citenamefont {Jacquemin}}]{loos2019reference}%
  \BibitemOpen
  \bibfield  {author} {\bibinfo {author} {\bibfnamefont {P.-F.}\ \bibnamefont
  {Loos}}, \bibinfo {author} {\bibfnamefont {M.}~\bibnamefont {Boggio-Pasqua}},
  \bibinfo {author} {\bibfnamefont {A.}~\bibnamefont {Scemama}}, \bibinfo
  {author} {\bibfnamefont {M.}~\bibnamefont {Caffarel}}, \ and\ \bibinfo
  {author} {\bibfnamefont {D.}~\bibnamefont {Jacquemin}},\ }\href {\doibase
  10.1021/acs.jctc.8b01205} {\bibfield  {journal} {\bibinfo  {journal} {J.
  Chem. Theory Comput.}\ }\textbf {\bibinfo {volume} {15}},\ \bibinfo {pages}
  {1939} (\bibinfo {year} {2019})},\ \Eprint
  {http://arxiv.org/abs/https://doi.org/10.1021/acs.jctc.8b01205}
  {https://doi.org/10.1021/acs.jctc.8b01205} \BibitemShut {NoStop}%
\bibitem [{\citenamefont {Lappe}\ and\ \citenamefont
  {Cave}(2000)}]{lappe2000on}%
  \BibitemOpen
  \bibfield  {author} {\bibinfo {author} {\bibfnamefont {J.}~\bibnamefont
  {Lappe}}\ and\ \bibinfo {author} {\bibfnamefont {R.~J.}\ \bibnamefont
  {Cave}},\ }\href {\doibase 10.1021/jp992518z} {\bibfield  {journal} {\bibinfo
   {journal} {J. Phys. Chem. A}\ }\textbf {\bibinfo {volume} {104}},\ \bibinfo
  {pages} {2294} (\bibinfo {year} {2000})},\ \Eprint
  {http://arxiv.org/abs/https://doi.org/10.1021/jp992518z}
  {https://doi.org/10.1021/jp992518z} \BibitemShut {NoStop}%
\bibitem [{\citenamefont {Serrano‐Andrés}\ \emph {et~al.}(1993)\citenamefont
  {Serrano‐Andrés}, \citenamefont {Merchán}, \citenamefont {Nebot‐Gil},
  \citenamefont {Lindh},\ and\ \citenamefont
  {Roos}}]{serranoandres1993towards}%
  \BibitemOpen
  \bibfield  {author} {\bibinfo {author} {\bibfnamefont {L.}~\bibnamefont
  {Serrano‐Andrés}}, \bibinfo {author} {\bibfnamefont {M.}~\bibnamefont
  {Merchán}}, \bibinfo {author} {\bibfnamefont {I.}~\bibnamefont
  {Nebot‐Gil}}, \bibinfo {author} {\bibfnamefont {R.}~\bibnamefont {Lindh}},
  \ and\ \bibinfo {author} {\bibfnamefont {B.~O.}\ \bibnamefont {Roos}},\
  }\href {\doibase 10.1063/1.465071} {\bibfield  {journal} {\bibinfo  {journal}
  {J. Chem. Phys.}\ }\textbf {\bibinfo {volume} {98}},\ \bibinfo {pages} {3151}
  (\bibinfo {year} {1993})},\ \Eprint
  {http://arxiv.org/abs/https://pubs.aip.org/aip/jcp/article-pdf/98/4/3151/15360596/3151\_1\_online.pdf}
  {https://pubs.aip.org/aip/jcp/article-pdf/98/4/3151/15360596/3151\_1\_online.pdf}
  \BibitemShut {NoStop}%
\bibitem [{\citenamefont {Hsu}\ \emph {et~al.}(2001)\citenamefont {Hsu},
  \citenamefont {Hirata},\ and\ \citenamefont
  {Head-Gordon}}]{hsu2001excitation}%
  \BibitemOpen
  \bibfield  {author} {\bibinfo {author} {\bibfnamefont {C.-P.}\ \bibnamefont
  {Hsu}}, \bibinfo {author} {\bibfnamefont {S.}~\bibnamefont {Hirata}}, \ and\
  \bibinfo {author} {\bibfnamefont {M.}~\bibnamefont {Head-Gordon}},\ }\href
  {\doibase 10.1021/jp0024367} {\bibfield  {journal} {\bibinfo  {journal} {J.
  Phys. Chem. A}\ }\textbf {\bibinfo {volume} {105}},\ \bibinfo {pages} {451}
  (\bibinfo {year} {2001})},\ \Eprint
  {http://arxiv.org/abs/https://doi.org/10.1021/jp0024367}
  {https://doi.org/10.1021/jp0024367} \BibitemShut {NoStop}%
\bibitem [{\citenamefont {Starcke}\ \emph {et~al.}(2006)\citenamefont
  {Starcke}, \citenamefont {Wormit}, \citenamefont {Schirmer},\ and\
  \citenamefont {Dreuw}}]{starcke2006how}%
  \BibitemOpen
  \bibfield  {author} {\bibinfo {author} {\bibfnamefont {J.~H.}\ \bibnamefont
  {Starcke}}, \bibinfo {author} {\bibfnamefont {M.}~\bibnamefont {Wormit}},
  \bibinfo {author} {\bibfnamefont {J.}~\bibnamefont {Schirmer}}, \ and\
  \bibinfo {author} {\bibfnamefont {A.}~\bibnamefont {Dreuw}},\ }\href
  {\doibase https://doi.org/10.1016/j.chemphys.2006.07.020} {\bibfield
  {journal} {\bibinfo  {journal} {Chem. Phys.}\ }\textbf {\bibinfo {volume}
  {329}},\ \bibinfo {pages} {39} (\bibinfo {year} {2006})},\ \bibinfo {note}
  {electron Correlation and Multimode Dynamics in Molecules}\BibitemShut
  {NoStop}%
\bibitem [{\citenamefont {Smith}\ and\ \citenamefont
  {Michl}(2010)}]{smith2010singlet}%
  \BibitemOpen
  \bibfield  {author} {\bibinfo {author} {\bibfnamefont {M.~B.}\ \bibnamefont
  {Smith}}\ and\ \bibinfo {author} {\bibfnamefont {J.}~\bibnamefont {Michl}},\
  }\href {\doibase 10.1021/cr1002613} {\bibfield  {journal} {\bibinfo
  {journal} {Chem. Rev.}\ }\textbf {\bibinfo {volume} {110}},\ \bibinfo {pages}
  {6891} (\bibinfo {year} {2010})},\ \bibinfo {note} {pMID: 21053979},\ \Eprint
  {http://arxiv.org/abs/https://doi.org/10.1021/cr1002613}
  {https://doi.org/10.1021/cr1002613} \BibitemShut {NoStop}%
\bibitem [{\citenamefont {Smith}\ and\ \citenamefont
  {Michl}(2013)}]{smith2013recent}%
  \BibitemOpen
  \bibfield  {author} {\bibinfo {author} {\bibfnamefont {M.~B.}\ \bibnamefont
  {Smith}}\ and\ \bibinfo {author} {\bibfnamefont {J.}~\bibnamefont {Michl}},\
  }\href {\doibase 10.1146/annurev-physchem-040412-110130} {\bibfield
  {journal} {\bibinfo  {journal} {Annu. Rev. Phys. Chem.}\ }\textbf {\bibinfo
  {volume} {64}},\ \bibinfo {pages} {361} (\bibinfo {year} {2013})},\ \bibinfo
  {note} {pMID: 23298243},\ \Eprint
  {http://arxiv.org/abs/https://doi.org/10.1146/annurev-physchem-040412-110130}
  {https://doi.org/10.1146/annurev-physchem-040412-110130} \BibitemShut
  {NoStop}%
\bibitem [{\citenamefont {Elliott}\ \emph {et~al.}(2011)\citenamefont
  {Elliott}, \citenamefont {Goldson}, \citenamefont {Canahui},\ and\
  \citenamefont {Maitra}}]{elliott2011perspectives}%
  \BibitemOpen
  \bibfield  {author} {\bibinfo {author} {\bibfnamefont {P.}~\bibnamefont
  {Elliott}}, \bibinfo {author} {\bibfnamefont {S.}~\bibnamefont {Goldson}},
  \bibinfo {author} {\bibfnamefont {C.}~\bibnamefont {Canahui}}, \ and\
  \bibinfo {author} {\bibfnamefont {N.~T.}\ \bibnamefont {Maitra}},\ }\href
  {\doibase https://doi.org/10.1016/j.chemphys.2011.03.020} {\bibfield
  {journal} {\bibinfo  {journal} {Chem. Phys.}\ }\textbf {\bibinfo {volume}
  {391}},\ \bibinfo {pages} {110} (\bibinfo {year} {2011})}\BibitemShut
  {NoStop}%
\bibitem [{\citenamefont {Runge}\ and\ \citenamefont
  {Gross}(1984)}]{runge1984density}%
  \BibitemOpen
  \bibfield  {author} {\bibinfo {author} {\bibfnamefont {E.}~\bibnamefont
  {Runge}}\ and\ \bibinfo {author} {\bibfnamefont {E.~K.~U.}\ \bibnamefont
  {Gross}},\ }\href@noop {} {\bibfield  {journal} {\bibinfo  {journal} {Phys.
  Rev. Lett.}\ }\textbf {\bibinfo {volume} {52}},\ \bibinfo {pages} {997}
  (\bibinfo {year} {1984})}\BibitemShut {NoStop}%
\bibitem [{\citenamefont {Casida}(1995)}]{casida1995timedependent}%
  \BibitemOpen
  \bibfield  {author} {\bibinfo {author} {\bibfnamefont {M.~E.}\ \bibnamefont
  {Casida}},\ }in\ \href@noop {} {\emph {\bibinfo {booktitle} {Recent Advances
  in Density Functional Methods}}},\ Vol.~\bibinfo {volume} {1},\ \bibinfo
  {editor} {edited by\ \bibinfo {editor} {\bibfnamefont {D.~P.}\ \bibnamefont
  {Chong}}}\ (\bibinfo  {publisher} {World Scientific},\ \bibinfo {year}
  {1995})\ pp.\ \bibinfo {pages} {155--192}\BibitemShut {NoStop}%
\bibitem [{\citenamefont {Casida}\ and\ \citenamefont
  {Huix-Rotllant}(2012)}]{casida2012progress}%
  \BibitemOpen
  \bibfield  {author} {\bibinfo {author} {\bibfnamefont {M.}~\bibnamefont
  {Casida}}\ and\ \bibinfo {author} {\bibfnamefont {M.}~\bibnamefont
  {Huix-Rotllant}},\ }\href {\doibase 10.1146/annurev-physchem-032511-143803}
  {\bibfield  {journal} {\bibinfo  {journal} {Annu. Rev. Phys. Chem.}\ }\textbf
  {\bibinfo {volume} {63}},\ \bibinfo {pages} {287} (\bibinfo {year} {2012})},\
  \Eprint
  {http://arxiv.org/abs/https://doi.org/10.1146/annurev-physchem-032511-143803}
  {https://doi.org/10.1146/annurev-physchem-032511-143803} \BibitemShut
  {NoStop}%
\bibitem [{\citenamefont {Lacombe}\ and\ \citenamefont
  {Maitra}(2023)}]{Lacombe2023_Non-adiabatic}%
  \BibitemOpen
  \bibfield  {author} {\bibinfo {author} {\bibfnamefont {L.}~\bibnamefont
  {Lacombe}}\ and\ \bibinfo {author} {\bibfnamefont {N.}~\bibnamefont
  {Maitra}},\ }\href {https://doi.org/10.1038/s41524-023-01061-0} {\bibfield
  {journal} {\bibinfo  {journal} {npj Comput Mater}\ }\textbf {\bibinfo
  {volume} {9}},\ \bibinfo {pages} {124} (\bibinfo {year} {2023})}\BibitemShut
  {NoStop}%
\bibitem [{\citenamefont {Maitra}\ \emph {et~al.}(2004)\citenamefont {Maitra},
  \citenamefont {Zhang}, \citenamefont {Cave},\ and\ \citenamefont
  {Burke}}]{maitra2004double}%
  \BibitemOpen
  \bibfield  {author} {\bibinfo {author} {\bibfnamefont {N.~T.}\ \bibnamefont
  {Maitra}}, \bibinfo {author} {\bibfnamefont {F.}~\bibnamefont {Zhang}},
  \bibinfo {author} {\bibfnamefont {R.~J.}\ \bibnamefont {Cave}}, \ and\
  \bibinfo {author} {\bibfnamefont {K.}~\bibnamefont {Burke}},\ }\href
  {\doibase 10.1063/1.1651060} {\bibfield  {journal} {\bibinfo  {journal} {J.
  Chem. Phys.}\ }\textbf {\bibinfo {volume} {120}},\ \bibinfo {pages} {5932}
  (\bibinfo {year} {2004})},\ \Eprint
  {http://arxiv.org/abs/https://pubs.aip.org/aip/jcp/article-pdf/120/13/5932/10855096/5932\_1\_online.pdf}
  {https://pubs.aip.org/aip/jcp/article-pdf/120/13/5932/10855096/5932\_1\_online.pdf}
  \BibitemShut {NoStop}%
\bibitem [{\citenamefont {Cave}\ \emph {et~al.}(2004)\citenamefont {Cave},
  \citenamefont {Zhang}, \citenamefont {Maitra},\ and\ \citenamefont
  {Burke}}]{cave2004dressed}%
  \BibitemOpen
  \bibfield  {author} {\bibinfo {author} {\bibfnamefont {R.~J.}\ \bibnamefont
  {Cave}}, \bibinfo {author} {\bibfnamefont {F.}~\bibnamefont {Zhang}},
  \bibinfo {author} {\bibfnamefont {N.~T.}\ \bibnamefont {Maitra}}, \ and\
  \bibinfo {author} {\bibfnamefont {K.}~\bibnamefont {Burke}},\ }\href
  {\doibase https://doi.org/10.1016/j.cplett.2004.03.051} {\bibfield  {journal}
  {\bibinfo  {journal} {Chem. Phys. Lett.}\ }\textbf {\bibinfo {volume}
  {389}},\ \bibinfo {pages} {39} (\bibinfo {year} {2004})}\BibitemShut
  {NoStop}%
\bibitem [{\citenamefont {Huix-Rotllant}\ \emph {et~al.}(2011)\citenamefont
  {Huix-Rotllant}, \citenamefont {Ipatov}, \citenamefont {Rubio},\ and\
  \citenamefont {Casida}}]{Huix-Rotllant2011_Assessment}%
  \BibitemOpen
  \bibfield  {author} {\bibinfo {author} {\bibfnamefont {M.}~\bibnamefont
  {Huix-Rotllant}}, \bibinfo {author} {\bibfnamefont {A.}~\bibnamefont
  {Ipatov}}, \bibinfo {author} {\bibfnamefont {A.}~\bibnamefont {Rubio}}, \
  and\ \bibinfo {author} {\bibfnamefont {M.~E.}\ \bibnamefont {Casida}},\
  }\href {\doibase https://doi.org/10.1016/j.chemphys.2011.03.019} {\bibfield
  {journal} {\bibinfo  {journal} {Chemical Physics}\ }\textbf {\bibinfo
  {volume} {391}},\ \bibinfo {pages} {120} (\bibinfo {year}
  {2011})}\BibitemShut {NoStop}%
\bibitem [{\citenamefont {Maitra}(2022)}]{maitra2022double}%
  \BibitemOpen
  \bibfield  {author} {\bibinfo {author} {\bibfnamefont {N.~T.}\ \bibnamefont
  {Maitra}},\ }\href {\doibase 10.1146/annurev-physchem-082720-124933}
  {\bibfield  {journal} {\bibinfo  {journal} {Annu. Rev. Phys. Chem.}\ }\textbf
  {\bibinfo {volume} {73}},\ \bibinfo {pages} {117} (\bibinfo {year} {2022})},\
  \Eprint
  {http://arxiv.org/abs/https://doi.org/10.1146/annurev-physchem-082720-124933}
  {https://doi.org/10.1146/annurev-physchem-082720-124933} \BibitemShut
  {NoStop}%
\bibitem [{\citenamefont {Fan}(1949)}]{Fan1949_On}%
  \BibitemOpen
  \bibfield  {author} {\bibinfo {author} {\bibfnamefont {K.}~\bibnamefont
  {Fan}},\ }\href {https://doi.org/10.1073/pnas.35.11.652} {\bibfield
  {journal} {\bibinfo  {journal} {PNAS}\ }\textbf {\bibinfo {volume} {35}},\
  \bibinfo {pages} {652} (\bibinfo {year} {1949})}\BibitemShut {NoStop}%
\bibitem [{\citenamefont {Theophilou}(1979)}]{theophilou1979energy}%
  \BibitemOpen
  \bibfield  {author} {\bibinfo {author} {\bibfnamefont {A.~K.}\ \bibnamefont
  {Theophilou}},\ }\href {https://doi.org/10.1088/0022-3719/12/24/013}
  {\bibfield  {journal} {\bibinfo  {journal} {J. Phys. C: Solid State Phys.}\
  }\textbf {\bibinfo {volume} {12}},\ \bibinfo {pages} {5419} (\bibinfo {year}
  {1979})}\BibitemShut {NoStop}%
\bibitem [{\citenamefont {Hendeković}(1982)}]{Hendekovic1982_equi-ensembles}%
  \BibitemOpen
  \bibfield  {author} {\bibinfo {author} {\bibfnamefont {J.}~\bibnamefont
  {Hendeković}},\ }\href {\doibase
  https://doi.org/10.1016/0009-2614(82)80024-9} {\bibfield  {journal} {\bibinfo
   {journal} {Chemical Physics Letters}\ }\textbf {\bibinfo {volume} {90}},\
  \bibinfo {pages} {198} (\bibinfo {year} {1982})}\BibitemShut {NoStop}%
\bibitem [{\citenamefont {Gross}\ \emph
  {et~al.}(1988{\natexlab{a}})\citenamefont {Gross}, \citenamefont {Oliveira},\
  and\ \citenamefont {Kohn}}]{gross1988rayleigh}%
  \BibitemOpen
  \bibfield  {author} {\bibinfo {author} {\bibfnamefont {E.~K.~U.}\
  \bibnamefont {Gross}}, \bibinfo {author} {\bibfnamefont {L.~N.}\ \bibnamefont
  {Oliveira}}, \ and\ \bibinfo {author} {\bibfnamefont {W.}~\bibnamefont
  {Kohn}},\ }\href {https://doi.org/10.1103/PhysRevA.37.2805} {\bibfield
  {journal} {\bibinfo  {journal} {Phys. Rev. A}\ }\textbf {\bibinfo {volume}
  {37}},\ \bibinfo {pages} {2805} (\bibinfo {year}
  {1988}{\natexlab{a}})}\BibitemShut {NoStop}%
\bibitem [{\citenamefont {Deur}\ \emph {et~al.}(2017)\citenamefont {Deur},
  \citenamefont {Mazouin},\ and\ \citenamefont {Fromager}}]{deur2017exact}%
  \BibitemOpen
  \bibfield  {author} {\bibinfo {author} {\bibfnamefont {K.}~\bibnamefont
  {Deur}}, \bibinfo {author} {\bibfnamefont {L.}~\bibnamefont {Mazouin}}, \
  and\ \bibinfo {author} {\bibfnamefont {E.}~\bibnamefont {Fromager}},\ }\href
  {https://doi.org/10.1103/PhysRevB.95.035120} {\bibfield  {journal} {\bibinfo
  {journal} {Phys. Rev. B}\ }\textbf {\bibinfo {volume} {95}},\ \bibinfo
  {pages} {035120} (\bibinfo {year} {2017})}\BibitemShut {NoStop}%
\bibitem [{\citenamefont {Yang}\ \emph {et~al.}(2017)\citenamefont {Yang},
  \citenamefont {Pribram-Jones}, \citenamefont {Burke},\ and\ \citenamefont
  {Ullrich}}]{yang2017direct}%
  \BibitemOpen
  \bibfield  {author} {\bibinfo {author} {\bibfnamefont {Z.-h.}\ \bibnamefont
  {Yang}}, \bibinfo {author} {\bibfnamefont {A.}~\bibnamefont {Pribram-Jones}},
  \bibinfo {author} {\bibfnamefont {K.}~\bibnamefont {Burke}}, \ and\ \bibinfo
  {author} {\bibfnamefont {C.~A.}\ \bibnamefont {Ullrich}},\ }\href
  {https://doi.org/10.1103/PhysRevLett.119.033003} {\bibfield  {journal}
  {\bibinfo  {journal} {Phys. Rev. Lett.}\ }\textbf {\bibinfo {volume} {119}},\
  \bibinfo {pages} {033003} (\bibinfo {year} {2017})}\BibitemShut {NoStop}%
\bibitem [{\citenamefont {Gould}\ and\ \citenamefont
  {Pittalis}(2017)}]{gould2017hartree}%
  \BibitemOpen
  \bibfield  {author} {\bibinfo {author} {\bibfnamefont {T.}~\bibnamefont
  {Gould}}\ and\ \bibinfo {author} {\bibfnamefont {S.}~\bibnamefont
  {Pittalis}},\ }\href {https://doi.org/10.1103/PhysRevLett.119.243001}
  {\bibfield  {journal} {\bibinfo  {journal} {Phys. Rev. Lett.}\ }\textbf
  {\bibinfo {volume} {119}},\ \bibinfo {pages} {243001} (\bibinfo {year}
  {2017})}\BibitemShut {NoStop}%
\bibitem [{\citenamefont {Gould}\ \emph {et~al.}(2018)\citenamefont {Gould},
  \citenamefont {Kronik},\ and\ \citenamefont {Pittalis}}]{gould2018charge}%
  \BibitemOpen
  \bibfield  {author} {\bibinfo {author} {\bibfnamefont {T.}~\bibnamefont
  {Gould}}, \bibinfo {author} {\bibfnamefont {L.}~\bibnamefont {Kronik}}, \
  and\ \bibinfo {author} {\bibfnamefont {S.}~\bibnamefont {Pittalis}},\ }\href
  {https://doi.org/10.1063/1.5022832} {\bibfield  {journal} {\bibinfo
  {journal} {J. Chem. Phys.}\ }\textbf {\bibinfo {volume} {148}},\ \bibinfo
  {pages} {174101} (\bibinfo {year} {2018})}\BibitemShut {NoStop}%
\bibitem [{\citenamefont {Deur}\ \emph {et~al.}(2018)\citenamefont {Deur},
  \citenamefont {Mazouin}, \citenamefont {Senjean},\ and\ \citenamefont
  {Fromager}}]{deur2018exploring}%
  \BibitemOpen
  \bibfield  {author} {\bibinfo {author} {\bibfnamefont {K.}~\bibnamefont
  {Deur}}, \bibinfo {author} {\bibfnamefont {L.}~\bibnamefont {Mazouin}},
  \bibinfo {author} {\bibfnamefont {B.}~\bibnamefont {Senjean}}, \ and\
  \bibinfo {author} {\bibfnamefont {E.}~\bibnamefont {Fromager}},\ }\href
  {https://doi.org/10.1140/epjb/e2018-90124-7} {\bibfield  {journal} {\bibinfo
  {journal} {Eur. Phys. J. B}\ }\textbf {\bibinfo {volume} {91}},\ \bibinfo
  {pages} {162} (\bibinfo {year} {2018})}\BibitemShut {NoStop}%
\bibitem [{\citenamefont {Gould}\ and\ \citenamefont
  {Pittalis}(2019)}]{PRL19_Gould_DD_correlation}%
  \BibitemOpen
  \bibfield  {author} {\bibinfo {author} {\bibfnamefont {T.}~\bibnamefont
  {Gould}}\ and\ \bibinfo {author} {\bibfnamefont {S.}~\bibnamefont
  {Pittalis}},\ }\href {\doibase 10.1103/PhysRevLett.123.016401} {\bibfield
  {journal} {\bibinfo  {journal} {Phys. Rev. Lett.}\ }\textbf {\bibinfo
  {volume} {123}},\ \bibinfo {pages} {016401} (\bibinfo {year}
  {2019})}\BibitemShut {NoStop}%
\bibitem [{\citenamefont {Fromager}(2020)}]{Fromager_2020}%
  \BibitemOpen
  \bibfield  {author} {\bibinfo {author} {\bibfnamefont {E.}~\bibnamefont
  {Fromager}},\ }\href {\doibase 10.1103/PhysRevLett.124.243001} {\bibfield
  {journal} {\bibinfo  {journal} {Phys. Rev. Lett.}\ }\textbf {\bibinfo
  {volume} {124}},\ \bibinfo {pages} {243001} (\bibinfo {year}
  {2020})}\BibitemShut {NoStop}%
\bibitem [{\citenamefont {Gould}\ \emph {et~al.}(2020)\citenamefont {Gould},
  \citenamefont {Stefanucci},\ and\ \citenamefont
  {Pittalis}}]{PRL20_Gould_Hartree_def_from_ACDF_th}%
  \BibitemOpen
  \bibfield  {author} {\bibinfo {author} {\bibfnamefont {T.}~\bibnamefont
  {Gould}}, \bibinfo {author} {\bibfnamefont {G.}~\bibnamefont {Stefanucci}}, \
  and\ \bibinfo {author} {\bibfnamefont {S.}~\bibnamefont {Pittalis}},\ }\href
  {\doibase 10.1103/PhysRevLett.125.233001} {\bibfield  {journal} {\bibinfo
  {journal} {Phys. Rev. Lett.}\ }\textbf {\bibinfo {volume} {125}},\ \bibinfo
  {pages} {233001} (\bibinfo {year} {2020})}\BibitemShut {NoStop}%
\bibitem [{\citenamefont {Gould}(2020)}]{Gould2020_Approximately}%
  \BibitemOpen
  \bibfield  {author} {\bibinfo {author} {\bibfnamefont {T.}~\bibnamefont
  {Gould}},\ }\href {\doibase 10.1021/acs.jpclett.0c02894} {\bibfield
  {journal} {\bibinfo  {journal} {J. Phys. Chem. Lett.}\ }\textbf {\bibinfo
  {volume} {11}},\ \bibinfo {pages} {9907} (\bibinfo {year}
  {2020})}\BibitemShut {NoStop}%
\bibitem [{\citenamefont {Loos}\ and\ \citenamefont
  {Fromager}(2020)}]{loos2020weight}%
  \BibitemOpen
  \bibfield  {author} {\bibinfo {author} {\bibfnamefont {P.-F.}\ \bibnamefont
  {Loos}}\ and\ \bibinfo {author} {\bibfnamefont {E.}~\bibnamefont
  {Fromager}},\ }\href {https://doi.org/10.1063/5.0007388} {\bibfield
  {journal} {\bibinfo  {journal} {J. Chem. Phys.}\ }\textbf {\bibinfo {volume}
  {152}},\ \bibinfo {pages} {214101} (\bibinfo {year} {2020})}\BibitemShut
  {NoStop}%
\bibitem [{\citenamefont {Gould}\ and\ \citenamefont
  {Kronik}(2021)}]{Gould2021_Ensemble_ugly}%
  \BibitemOpen
  \bibfield  {author} {\bibinfo {author} {\bibfnamefont {T.}~\bibnamefont
  {Gould}}\ and\ \bibinfo {author} {\bibfnamefont {L.}~\bibnamefont {Kronik}},\
  }\href {\doibase 10.1063/5.0040447} {\bibfield  {journal} {\bibinfo
  {journal} {J. Chem. Phys.}\ }\textbf {\bibinfo {volume} {154}},\ \bibinfo
  {pages} {094125} (\bibinfo {year} {2021})}\BibitemShut {NoStop}%
\bibitem [{\citenamefont {Gould}\ and\ \citenamefont
  {Pittalis}(2023)}]{gould2023local}%
  \BibitemOpen
  \bibfield  {author} {\bibinfo {author} {\bibfnamefont {T.}~\bibnamefont
  {Gould}}\ and\ \bibinfo {author} {\bibfnamefont {S.}~\bibnamefont
  {Pittalis}},\ }\href {https://arxiv.org/abs/2306.04023} {\enquote {\bibinfo
  {title} {Local density approximation for excited states},}\ } (\bibinfo
  {year} {2023}),\ \Eprint {http://arxiv.org/abs/2306.04023} {arXiv:2306.04023
  [physics.chem-ph]} \BibitemShut {NoStop}%
\bibitem [{\citenamefont {Gould}\ \emph {et~al.}(2023)\citenamefont {Gould},
  \citenamefont {Kooi}, \citenamefont {Gori-Giorgi},\ and\ \citenamefont
  {Pittalis}}]{Gould2023_Electronic}%
  \BibitemOpen
  \bibfield  {author} {\bibinfo {author} {\bibfnamefont {T.}~\bibnamefont
  {Gould}}, \bibinfo {author} {\bibfnamefont {D.~P.}\ \bibnamefont {Kooi}},
  \bibinfo {author} {\bibfnamefont {P.}~\bibnamefont {Gori-Giorgi}}, \ and\
  \bibinfo {author} {\bibfnamefont {S.}~\bibnamefont {Pittalis}},\ }\href
  {\doibase 10.1103/PhysRevLett.130.106401} {\bibfield  {journal} {\bibinfo
  {journal} {Phys. Rev. Lett.}\ }\textbf {\bibinfo {volume} {130}},\ \bibinfo
  {pages} {106401} (\bibinfo {year} {2023})}\BibitemShut {NoStop}%
\bibitem [{\citenamefont {Gould}\ \emph {et~al.}(2022)\citenamefont {Gould},
  \citenamefont {Hashimi}, \citenamefont {Kronik},\ and\ \citenamefont
  {Dale}}]{gould2022single}%
  \BibitemOpen
  \bibfield  {author} {\bibinfo {author} {\bibfnamefont {T.}~\bibnamefont
  {Gould}}, \bibinfo {author} {\bibfnamefont {Z.}~\bibnamefont {Hashimi}},
  \bibinfo {author} {\bibfnamefont {L.}~\bibnamefont {Kronik}}, \ and\ \bibinfo
  {author} {\bibfnamefont {S.~G.}\ \bibnamefont {Dale}},\ }\href {\doibase
  10.1021/acs.jpclett.2c00042} {\bibfield  {journal} {\bibinfo  {journal} {J.
  Phys. Chem. Lett.}\ }\textbf {\bibinfo {volume} {13}},\ \bibinfo {pages}
  {2452} (\bibinfo {year} {2022})},\ \Eprint
  {http://arxiv.org/abs/https://doi.org/10.1021/acs.jpclett.2c00042}
  {https://doi.org/10.1021/acs.jpclett.2c00042} \BibitemShut {NoStop}%
\bibitem [{\citenamefont {Gould}\ \emph {et~al.}(2021)\citenamefont {Gould},
  \citenamefont {Kronik},\ and\ \citenamefont {Pittalis}}]{gould2021double}%
  \BibitemOpen
  \bibfield  {author} {\bibinfo {author} {\bibfnamefont {T.}~\bibnamefont
  {Gould}}, \bibinfo {author} {\bibfnamefont {L.}~\bibnamefont {Kronik}}, \
  and\ \bibinfo {author} {\bibfnamefont {S.}~\bibnamefont {Pittalis}},\ }\href
  {\doibase 10.1103/PhysRevA.104.022803} {\bibfield  {journal} {\bibinfo
  {journal} {Phys. Rev. A}\ }\textbf {\bibinfo {volume} {104}},\ \bibinfo
  {pages} {022803} (\bibinfo {year} {2021})}\BibitemShut {NoStop}%
\bibitem [{\citenamefont {Cernatic}\ \emph {et~al.}(2022)\citenamefont
  {Cernatic}, \citenamefont {Senjean}, \citenamefont {Robert},\ and\
  \citenamefont {Fromager}}]{Cernatic2022}%
  \BibitemOpen
  \bibfield  {author} {\bibinfo {author} {\bibfnamefont {F.}~\bibnamefont
  {Cernatic}}, \bibinfo {author} {\bibfnamefont {B.}~\bibnamefont {Senjean}},
  \bibinfo {author} {\bibfnamefont {V.}~\bibnamefont {Robert}}, \ and\ \bibinfo
  {author} {\bibfnamefont {E.}~\bibnamefont {Fromager}},\ }\href
  {https://doi.org/10.1007/s41061-021-00359-1} {\bibfield  {journal} {\bibinfo
  {journal} {Top Curr Chem (Z)}\ }\textbf {\bibinfo {volume} {380}},\ \bibinfo
  {pages} {4} (\bibinfo {year} {2022})}\BibitemShut {NoStop}%
\bibitem [{\citenamefont {Schilling}\ and\ \citenamefont
  {Pittalis}(2021)}]{Schilling2021_Ensemble}%
  \BibitemOpen
  \bibfield  {author} {\bibinfo {author} {\bibfnamefont {C.}~\bibnamefont
  {Schilling}}\ and\ \bibinfo {author} {\bibfnamefont {S.}~\bibnamefont
  {Pittalis}},\ }\href {\doibase 10.1103/PhysRevLett.127.023001} {\bibfield
  {journal} {\bibinfo  {journal} {Phys. Rev. Lett.}\ }\textbf {\bibinfo
  {volume} {127}},\ \bibinfo {pages} {023001} (\bibinfo {year}
  {2021})}\BibitemShut {NoStop}%
\bibitem [{\citenamefont {Liebert}\ \emph {et~al.}(2022)\citenamefont
  {Liebert}, \citenamefont {Castillo}, \citenamefont {Labb{\'e}},\ and\
  \citenamefont {Schilling}}]{Liebert2022_Foundation}%
  \BibitemOpen
  \bibfield  {author} {\bibinfo {author} {\bibfnamefont {J.}~\bibnamefont
  {Liebert}}, \bibinfo {author} {\bibfnamefont {F.}~\bibnamefont {Castillo}},
  \bibinfo {author} {\bibfnamefont {J.-P.}\ \bibnamefont {Labb{\'e}}}, \ and\
  \bibinfo {author} {\bibfnamefont {C.}~\bibnamefont {Schilling}},\ }\href
  {\doibase 10.1021/acs.jctc.1c00561} {\bibfield  {journal} {\bibinfo
  {journal} {J. Chem. Theory Comput.}\ }\textbf {\bibinfo {volume} {18}},\
  \bibinfo {pages} {124} (\bibinfo {year} {2022})}\BibitemShut {NoStop}%
\bibitem [{\citenamefont {Benavides-Riveros}\ \emph {et~al.}(2022)\citenamefont
  {Benavides-Riveros}, \citenamefont {Chen}, \citenamefont {Schilling},
  \citenamefont {Mantilla},\ and\ \citenamefont
  {Pittalis}}]{Benavides-Riveros2022_Excitations}%
  \BibitemOpen
  \bibfield  {author} {\bibinfo {author} {\bibfnamefont {C.~L.}\ \bibnamefont
  {Benavides-Riveros}}, \bibinfo {author} {\bibfnamefont {L.}~\bibnamefont
  {Chen}}, \bibinfo {author} {\bibfnamefont {C.}~\bibnamefont {Schilling}},
  \bibinfo {author} {\bibfnamefont {S.}~\bibnamefont {Mantilla}}, \ and\
  \bibinfo {author} {\bibfnamefont {S.}~\bibnamefont {Pittalis}},\ }\href
  {\doibase 10.1103/PhysRevLett.129.066401} {\bibfield  {journal} {\bibinfo
  {journal} {Phys. Rev. Lett.}\ }\textbf {\bibinfo {volume} {129}},\ \bibinfo
  {pages} {066401} (\bibinfo {year} {2022})}\BibitemShut {NoStop}%
\bibitem [{\citenamefont {Liebert}\ and\ \citenamefont
  {Schilling}(2023{\natexlab{a}})}]{Liebert_2023_An_exact_bosons}%
  \BibitemOpen
  \bibfield  {author} {\bibinfo {author} {\bibfnamefont {J.}~\bibnamefont
  {Liebert}}\ and\ \bibinfo {author} {\bibfnamefont {C.}~\bibnamefont
  {Schilling}},\ }\href {\doibase 10.1088/1367-2630/acb006} {\bibfield
  {journal} {\bibinfo  {journal} {New J. Phys.}\ }\textbf {\bibinfo {volume}
  {25}},\ \bibinfo {pages} {013009} (\bibinfo {year}
  {2023}{\natexlab{a}})}\BibitemShut {NoStop}%
\bibitem [{\citenamefont {Liebert}\ and\ \citenamefont
  {Schilling}(2023{\natexlab{b}})}]{Liebert2023_Deriving}%
  \BibitemOpen
  \bibfield  {author} {\bibinfo {author} {\bibfnamefont {J.}~\bibnamefont
  {Liebert}}\ and\ \bibinfo {author} {\bibfnamefont {C.}~\bibnamefont
  {Schilling}},\ }\href {\doibase 10.21468/SciPostPhys.14.5.120} {\bibfield
  {journal} {\bibinfo  {journal} {SciPost Phys.}\ }\textbf {\bibinfo {volume}
  {14}},\ \bibinfo {pages} {120} (\bibinfo {year}
  {2023}{\natexlab{b}})}\BibitemShut {NoStop}%
\bibitem [{\citenamefont {Ding}\ \emph {et~al.}(2024)\citenamefont {Ding},
  \citenamefont {Hong},\ and\ \citenamefont {Schilling}}]{ding2024ground}%
  \BibitemOpen
  \bibfield  {author} {\bibinfo {author} {\bibfnamefont {L.}~\bibnamefont
  {Ding}}, \bibinfo {author} {\bibfnamefont {C.-L.}\ \bibnamefont {Hong}}, \
  and\ \bibinfo {author} {\bibfnamefont {C.}~\bibnamefont {Schilling}},\ }\href
  {https://arxiv.org/abs/2401.12104} {\enquote {\bibinfo {title} {Ground and
  excited states from ensemble variational principles},}\ } (\bibinfo {year}
  {2024}),\ \Eprint {http://arxiv.org/abs/2401.12104} {arXiv:2401.12104
  [quant-ph]} \BibitemShut {NoStop}%
\bibitem [{\citenamefont {Theophilou}(1987)}]{theophilou_book}%
  \BibitemOpen
  \bibfield  {author} {\bibinfo {author} {\bibfnamefont {A.~K.}\ \bibnamefont
  {Theophilou}},\ }\enquote {\bibinfo {title} {The single particle density in
  physics and chemistry},}\ \ (\bibinfo  {publisher} {Academic Press},\
  \bibinfo {year} {1987})\ pp.\ \bibinfo {pages} {210--212}\BibitemShut
  {NoStop}%
\bibitem [{\citenamefont {Gross}\ \emph
  {et~al.}(1988{\natexlab{b}})\citenamefont {Gross}, \citenamefont {Oliveira},\
  and\ \citenamefont {Kohn}}]{gross1988density}%
  \BibitemOpen
  \bibfield  {author} {\bibinfo {author} {\bibfnamefont {E.~K.~U.}\
  \bibnamefont {Gross}}, \bibinfo {author} {\bibfnamefont {L.~N.}\ \bibnamefont
  {Oliveira}}, \ and\ \bibinfo {author} {\bibfnamefont {W.}~\bibnamefont
  {Kohn}},\ }\href {https://doi.org/10.1103/PhysRevA.37.2809} {\bibfield
  {journal} {\bibinfo  {journal} {Phys. Rev. A}\ }\textbf {\bibinfo {volume}
  {37}},\ \bibinfo {pages} {2809} (\bibinfo {year}
  {1988}{\natexlab{b}})}\BibitemShut {NoStop}%
\bibitem [{\citenamefont {Oliveira}\ \emph {et~al.}(1988)\citenamefont
  {Oliveira}, \citenamefont {Gross},\ and\ \citenamefont
  {Kohn}}]{oliveira1988density}%
  \BibitemOpen
  \bibfield  {author} {\bibinfo {author} {\bibfnamefont {L.~N.}\ \bibnamefont
  {Oliveira}}, \bibinfo {author} {\bibfnamefont {E.~K.~U.}\ \bibnamefont
  {Gross}}, \ and\ \bibinfo {author} {\bibfnamefont {W.}~\bibnamefont {Kohn}},\
  }\href@noop {} {\bibfield  {journal} {\bibinfo  {journal} {Phys. Rev. A}\
  }\textbf {\bibinfo {volume} {37}},\ \bibinfo {pages} {2821} (\bibinfo {year}
  {1988})}\BibitemShut {NoStop}%
\bibitem [{\citenamefont {Cernatic}\ \emph {et~al.}(2024)\citenamefont
  {Cernatic}, \citenamefont {Loos}, \citenamefont {Senjean},\ and\
  \citenamefont {Fromager}}]{cernatic2024neutral}%
  \BibitemOpen
  \bibfield  {author} {\bibinfo {author} {\bibfnamefont {F.}~\bibnamefont
  {Cernatic}}, \bibinfo {author} {\bibfnamefont {P.-F.}\ \bibnamefont {Loos}},
  \bibinfo {author} {\bibfnamefont {B.}~\bibnamefont {Senjean}}, \ and\
  \bibinfo {author} {\bibfnamefont {E.}~\bibnamefont {Fromager}},\ }\href
  {https://arxiv.org/abs/2401.04685} {\enquote {\bibinfo {title} {Neutral
  electronic excitations and derivative discontinuities: An extended
  $n$-centered ensemble density functional theory perspective},}\ } (\bibinfo
  {year} {2024}),\ \Eprint {http://arxiv.org/abs/2401.04685} {arXiv:2401.04685
  [physics.chem-ph]} \BibitemShut {NoStop}%
\bibitem [{\citenamefont {Deur}\ and\ \citenamefont
  {Fromager}(2019)}]{deur2019ground}%
  \BibitemOpen
  \bibfield  {author} {\bibinfo {author} {\bibfnamefont {K.}~\bibnamefont
  {Deur}}\ and\ \bibinfo {author} {\bibfnamefont {E.}~\bibnamefont
  {Fromager}},\ }\href {https://doi.org/10.1063/1.5084312} {\bibfield
  {journal} {\bibinfo  {journal} {J. Chem. Phys.}\ }\textbf {\bibinfo {volume}
  {150}},\ \bibinfo {pages} {094106} (\bibinfo {year} {2019})}\BibitemShut
  {NoStop}%
\bibitem [{\citenamefont {Sagredo}\ and\ \citenamefont
  {Burke}(2018)}]{sagredo2018can}%
  \BibitemOpen
  \bibfield  {author} {\bibinfo {author} {\bibfnamefont {F.}~\bibnamefont
  {Sagredo}}\ and\ \bibinfo {author} {\bibfnamefont {K.}~\bibnamefont
  {Burke}},\ }\href {https://doi.org/10.1063/1.5043411} {\bibfield  {journal}
  {\bibinfo  {journal} {J. Chem. Phys.}\ }\textbf {\bibinfo {volume} {149}},\
  \bibinfo {pages} {134103} (\bibinfo {year} {2018})}\BibitemShut {NoStop}%
\bibitem [{\citenamefont {Marut}\ \emph {et~al.}(2020)\citenamefont {Marut},
  \citenamefont {Senjean}, \citenamefont {Fromager},\ and\ \citenamefont
  {Loos}}]{marut2020weight}%
  \BibitemOpen
  \bibfield  {author} {\bibinfo {author} {\bibfnamefont {C.}~\bibnamefont
  {Marut}}, \bibinfo {author} {\bibfnamefont {B.}~\bibnamefont {Senjean}},
  \bibinfo {author} {\bibfnamefont {E.}~\bibnamefont {Fromager}}, \ and\
  \bibinfo {author} {\bibfnamefont {P.-F.}\ \bibnamefont {Loos}},\ }\href
  {https://doi.org/10.1039/D0FD00059K} {\bibfield  {journal} {\bibinfo
  {journal} {Faraday Discuss.}\ }\textbf {\bibinfo {volume} {224}},\ \bibinfo
  {pages} {402} (\bibinfo {year} {2020})}\BibitemShut {NoStop}%
\bibitem [{\citenamefont {Yang}(2021)}]{Yang2021_Second}%
  \BibitemOpen
  \bibfield  {author} {\bibinfo {author} {\bibfnamefont {Z.-h.}\ \bibnamefont
  {Yang}},\ }\href {\doibase 10.1103/PhysRevA.104.052806} {\bibfield  {journal}
  {\bibinfo  {journal} {Phys. Rev. A}\ }\textbf {\bibinfo {volume} {104}},\
  \bibinfo {pages} {052806} (\bibinfo {year} {2021})}\BibitemShut {NoStop}%
\bibitem [{\citenamefont {Levy}(1995)}]{levy1995excitation}%
  \BibitemOpen
  \bibfield  {author} {\bibinfo {author} {\bibfnamefont {M.}~\bibnamefont
  {Levy}},\ }\href {https://doi.org/10.1103/PhysRevA.52.R4313} {\bibfield
  {journal} {\bibinfo  {journal} {Phys. Rev. A}\ }\textbf {\bibinfo {volume}
  {52}},\ \bibinfo {pages} {R4313} (\bibinfo {year} {1995})}\BibitemShut
  {NoStop}%
\bibitem [{\citenamefont {Senjean}\ and\ \citenamefont
  {Fromager}(2018)}]{senjean2018unified}%
  \BibitemOpen
  \bibfield  {author} {\bibinfo {author} {\bibfnamefont {B.}~\bibnamefont
  {Senjean}}\ and\ \bibinfo {author} {\bibfnamefont {E.}~\bibnamefont
  {Fromager}},\ }\href {https://doi.org/10.1103/PhysRevA.98.022513} {\bibfield
  {journal} {\bibinfo  {journal} {Phys. Rev. A}\ }\textbf {\bibinfo {volume}
  {98}},\ \bibinfo {pages} {022513} (\bibinfo {year} {2018})}\BibitemShut
  {NoStop}%
\bibitem [{\citenamefont {Senjean}\ and\ \citenamefont
  {Fromager}(2020)}]{senjean2020n}%
  \BibitemOpen
  \bibfield  {author} {\bibinfo {author} {\bibfnamefont {B.}~\bibnamefont
  {Senjean}}\ and\ \bibinfo {author} {\bibfnamefont {E.}~\bibnamefont
  {Fromager}},\ }\href {https://doi.org/10.1002/qua.26190} {\bibfield
  {journal} {\bibinfo  {journal} {Int. J. Quantum Chem.}\ }\textbf {\bibinfo
  {volume} {120}},\ \bibinfo {pages} {e26190} (\bibinfo {year}
  {2020})}\BibitemShut {NoStop}%
\bibitem [{\citenamefont {Perdew}\ \emph {et~al.}(1982)\citenamefont {Perdew},
  \citenamefont {Parr}, \citenamefont {Levy},\ and\ \citenamefont
  {Balduz~Jr}}]{perdew1982density}%
  \BibitemOpen
  \bibfield  {author} {\bibinfo {author} {\bibfnamefont {J.~P.}\ \bibnamefont
  {Perdew}}, \bibinfo {author} {\bibfnamefont {R.~G.}\ \bibnamefont {Parr}},
  \bibinfo {author} {\bibfnamefont {M.}~\bibnamefont {Levy}}, \ and\ \bibinfo
  {author} {\bibfnamefont {J.~L.}\ \bibnamefont {Balduz~Jr}},\ }\href
  {https://doi.org/10.1103/PhysRevLett.49.1691} {\bibfield  {journal} {\bibinfo
   {journal} {Phys. Rev. Lett.}\ }\textbf {\bibinfo {volume} {49}},\ \bibinfo
  {pages} {1691} (\bibinfo {year} {1982})}\BibitemShut {NoStop}%
\bibitem [{\citenamefont {Perdew}\ and\ \citenamefont
  {Levy}(1983)}]{perdew1983physical}%
  \BibitemOpen
  \bibfield  {author} {\bibinfo {author} {\bibfnamefont {J.~P.}\ \bibnamefont
  {Perdew}}\ and\ \bibinfo {author} {\bibfnamefont {M.}~\bibnamefont {Levy}},\
  }\href {https://doi.org/10.1103/PhysRevLett.51.1884} {\bibfield  {journal}
  {\bibinfo  {journal} {Phys. Rev. Lett.}\ }\textbf {\bibinfo {volume} {51}},\
  \bibinfo {pages} {1884} (\bibinfo {year} {1983})}\BibitemShut {NoStop}%
\bibitem [{\citenamefont {Baerends}\ \emph {et~al.}(2013)\citenamefont
  {Baerends}, \citenamefont {Gritsenko},\ and\ \citenamefont
  {Van~Meer}}]{baerends2013kohn}%
  \BibitemOpen
  \bibfield  {author} {\bibinfo {author} {\bibfnamefont {E.~J.}\ \bibnamefont
  {Baerends}}, \bibinfo {author} {\bibfnamefont {O.~V.}\ \bibnamefont
  {Gritsenko}}, \ and\ \bibinfo {author} {\bibfnamefont {R.}~\bibnamefont
  {Van~Meer}},\ }\href {http://dx.doi.org/10.1039/C3CP52547C} {\bibfield
  {journal} {\bibinfo  {journal} {Phys. Chem. Chem. Phys.}\ }\textbf {\bibinfo
  {volume} {15}},\ \bibinfo {pages} {16408} (\bibinfo {year}
  {2013})}\BibitemShut {NoStop}%
\bibitem [{\citenamefont {Baerends}(2017)}]{baerends2017kohn}%
  \BibitemOpen
  \bibfield  {author} {\bibinfo {author} {\bibfnamefont {E.~J.}\ \bibnamefont
  {Baerends}},\ }\href {http://dx.doi.org/10.1039/C7CP02123B} {\bibfield
  {journal} {\bibinfo  {journal} {Phys. Chem. Chem. Phys.}\ }\textbf {\bibinfo
  {volume} {19}},\ \bibinfo {pages} {15639} (\bibinfo {year}
  {2017})}\BibitemShut {NoStop}%
\bibitem [{\citenamefont {Baerends}(2018)}]{baerends2018density}%
  \BibitemOpen
  \bibfield  {author} {\bibinfo {author} {\bibfnamefont {E.~J.}\ \bibnamefont
  {Baerends}},\ }\href {https://doi.org/10.1063/1.5026951} {\bibfield
  {journal} {\bibinfo  {journal} {J. Chem. Phys.}\ }\textbf {\bibinfo {volume}
  {149}},\ \bibinfo {pages} {054105} (\bibinfo {year} {2018})}\BibitemShut
  {NoStop}%
\bibitem [{\citenamefont {Baerends}(2022)}]{Baerends2022_Chemical}%
  \BibitemOpen
  \bibfield  {author} {\bibinfo {author} {\bibfnamefont {E.~J.}\ \bibnamefont
  {Baerends}},\ }\href {\doibase 10.1039/D2CP01585D} {\bibfield  {journal}
  {\bibinfo  {journal} {Phys. Chem. Chem. Phys.}\ }\textbf {\bibinfo {volume}
  {24}},\ \bibinfo {pages} {12745} (\bibinfo {year} {2022})}\BibitemShut
  {NoStop}%
\bibitem [{\citenamefont {Baerends}(2020)}]{Baerends2020_On_derivatives}%
  \BibitemOpen
  \bibfield  {author} {\bibinfo {author} {\bibfnamefont {E.~J.}\ \bibnamefont
  {Baerends}},\ }\href {\doibase 10.1080/00268976.2019.1612955} {\bibfield
  {journal} {\bibinfo  {journal} {Mol. Phys.}\ }\textbf {\bibinfo {volume}
  {118}},\ \bibinfo {pages} {e1612955} (\bibinfo {year} {2020})}\BibitemShut
  {NoStop}%
\bibitem [{\citenamefont {Levy}(1979)}]{levy1979universal}%
  \BibitemOpen
  \bibfield  {author} {\bibinfo {author} {\bibfnamefont {M.}~\bibnamefont
  {Levy}},\ }\href {https://doi.org/10.1073/pnas.76.12.6062} {\bibfield
  {journal} {\bibinfo  {journal} {Proc. Natl. Acad. Sci.}\ }\textbf {\bibinfo
  {volume} {76}},\ \bibinfo {pages} {6062} (\bibinfo {year}
  {1979})}\BibitemShut {NoStop}%
\bibitem [{\citenamefont {Hodgson}\ \emph {et~al.}(2021)\citenamefont
  {Hodgson}, \citenamefont {Wetherell},\ and\ \citenamefont
  {Fromager}}]{PRA21_Hodgson_exact_Nc-eDFT_1D}%
  \BibitemOpen
  \bibfield  {author} {\bibinfo {author} {\bibfnamefont {M.~J.~P.}\
  \bibnamefont {Hodgson}}, \bibinfo {author} {\bibfnamefont {J.}~\bibnamefont
  {Wetherell}}, \ and\ \bibinfo {author} {\bibfnamefont {E.}~\bibnamefont
  {Fromager}},\ }\href {\doibase 10.1103/PhysRevA.103.012806} {\bibfield
  {journal} {\bibinfo  {journal} {Phys. Rev. A}\ }\textbf {\bibinfo {volume}
  {103}},\ \bibinfo {pages} {012806} (\bibinfo {year} {2021})}\BibitemShut
  {NoStop}%
\bibitem [{\citenamefont {Li}\ \emph {et~al.}(2018)\citenamefont {Li},
  \citenamefont {Requist},\ and\ \citenamefont {Gross}}]{li2018density}%
  \BibitemOpen
  \bibfield  {author} {\bibinfo {author} {\bibfnamefont {C.}~\bibnamefont
  {Li}}, \bibinfo {author} {\bibfnamefont {R.}~\bibnamefont {Requist}}, \ and\
  \bibinfo {author} {\bibfnamefont {E.~K.~U.}\ \bibnamefont {Gross}},\ }\href
  {https://doi.org/10.1063/1.5011663} {\bibfield  {journal} {\bibinfo
  {journal} {J. Chem. Phys.}\ }\textbf {\bibinfo {volume} {148}},\ \bibinfo
  {pages} {084110} (\bibinfo {year} {2018})}\BibitemShut {NoStop}%
\bibitem [{\citenamefont {Carrascal}\ \emph {et~al.}(2015)\citenamefont
  {Carrascal}, \citenamefont {Ferrer}, \citenamefont {Smith},\ and\
  \citenamefont {Burke}}]{carrascal2015hubbard}%
  \BibitemOpen
  \bibfield  {author} {\bibinfo {author} {\bibfnamefont {D.~J.}\ \bibnamefont
  {Carrascal}}, \bibinfo {author} {\bibfnamefont {J.}~\bibnamefont {Ferrer}},
  \bibinfo {author} {\bibfnamefont {J.~C.}\ \bibnamefont {Smith}}, \ and\
  \bibinfo {author} {\bibfnamefont {K.}~\bibnamefont {Burke}},\ }\href
  {http://stacks.iop.org/0953-8984/27/i=39/a=393001} {\bibfield  {journal}
  {\bibinfo  {journal} {J. Phys. Condens. Matter}\ }\textbf {\bibinfo {volume}
  {27}},\ \bibinfo {pages} {393001} (\bibinfo {year} {2015})}\BibitemShut
  {NoStop}%
\bibitem [{\citenamefont {Carrascal}\ \emph {et~al.}(2018)\citenamefont
  {Carrascal}, \citenamefont {Ferrer}, \citenamefont {Maitra},\ and\
  \citenamefont {Burke}}]{carrascal2018linear}%
  \BibitemOpen
  \bibfield  {author} {\bibinfo {author} {\bibfnamefont {D.~J.}\ \bibnamefont
  {Carrascal}}, \bibinfo {author} {\bibfnamefont {J.}~\bibnamefont {Ferrer}},
  \bibinfo {author} {\bibfnamefont {N.}~\bibnamefont {Maitra}}, \ and\ \bibinfo
  {author} {\bibfnamefont {K.}~\bibnamefont {Burke}},\ }\href
  {https://doi.org/10.1140/epjb/e2018-90114-9} {\bibfield  {journal} {\bibinfo
  {journal} {Eur. Phys. J. B}\ }\textbf {\bibinfo {volume} {91}},\ \bibinfo
  {pages} {142} (\bibinfo {year} {2018})}\BibitemShut {NoStop}%
\bibitem [{\citenamefont {Smith}\ \emph {et~al.}(2016)\citenamefont {Smith},
  \citenamefont {Pribram-Jones},\ and\ \citenamefont {Burke}}]{smith2016exact}%
  \BibitemOpen
  \bibfield  {author} {\bibinfo {author} {\bibfnamefont {J.~C.}\ \bibnamefont
  {Smith}}, \bibinfo {author} {\bibfnamefont {A.}~\bibnamefont
  {Pribram-Jones}}, \ and\ \bibinfo {author} {\bibfnamefont {K.}~\bibnamefont
  {Burke}},\ }\href {\doibase 10.1103/PhysRevB.93.245131} {\bibfield  {journal}
  {\bibinfo  {journal} {Phys. Rev. B}\ }\textbf {\bibinfo {volume} {93}},\
  \bibinfo {pages} {245131} (\bibinfo {year} {2016})}\BibitemShut {NoStop}%
\bibitem [{\citenamefont {Giarrusso}\ and\ \citenamefont
  {Loos}(2023)}]{Giarrusso2023_Exact}%
  \BibitemOpen
  \bibfield  {author} {\bibinfo {author} {\bibfnamefont {S.}~\bibnamefont
  {Giarrusso}}\ and\ \bibinfo {author} {\bibfnamefont {P.-F.}\ \bibnamefont
  {Loos}},\ }\href {\doibase 10.1021/acs.jpclett.3c02052} {\bibfield  {journal}
  {\bibinfo  {journal} {J. Phys. Chem. Lett.}\ }\textbf {\bibinfo {volume}
  {14}},\ \bibinfo {pages} {8780} (\bibinfo {year} {2023})}\BibitemShut
  {NoStop}%
\bibitem [{\citenamefont {Ullrich}(2018)}]{Ullrich2018_Density}%
  \BibitemOpen
  \bibfield  {author} {\bibinfo {author} {\bibfnamefont {C.~A.}\ \bibnamefont
  {Ullrich}},\ }\href {\doibase 10.1103/PhysRevB.98.035140} {\bibfield
  {journal} {\bibinfo  {journal} {Phys. Rev. B}\ }\textbf {\bibinfo {volume}
  {98}},\ \bibinfo {pages} {035140} (\bibinfo {year} {2018})}\BibitemShut
  {NoStop}%
\bibitem [{\citenamefont {Scott}\ \emph {et~al.}(2023)\citenamefont {Scott},
  \citenamefont {Kozlowski}, \citenamefont {Crisostomo}, \citenamefont
  {Pribram-Jones},\ and\ \citenamefont {Burke}}]{scott2023exact}%
  \BibitemOpen
  \bibfield  {author} {\bibinfo {author} {\bibfnamefont {T.~R.}\ \bibnamefont
  {Scott}}, \bibinfo {author} {\bibfnamefont {J.}~\bibnamefont {Kozlowski}},
  \bibinfo {author} {\bibfnamefont {S.}~\bibnamefont {Crisostomo}}, \bibinfo
  {author} {\bibfnamefont {A.}~\bibnamefont {Pribram-Jones}}, \ and\ \bibinfo
  {author} {\bibfnamefont {K.}~\bibnamefont {Burke}},\ }\href@noop {} {\enquote
  {\bibinfo {title} {Exact conditions for ensemble density functional
  theory},}\ } (\bibinfo {year} {2023}),\ \Eprint
  {http://arxiv.org/abs/2307.00187} {arXiv:2307.00187 [cond-mat.str-el]}
  \BibitemShut {NoStop}%
\bibitem [{\citenamefont {Sobrino}\ \emph {et~al.}(2023)\citenamefont
  {Sobrino}, \citenamefont {Jacob},\ and\ \citenamefont
  {Kurth}}]{Sobrino2023_What}%
  \BibitemOpen
  \bibfield  {author} {\bibinfo {author} {\bibfnamefont {N.}~\bibnamefont
  {Sobrino}}, \bibinfo {author} {\bibfnamefont {D.}~\bibnamefont {Jacob}}, \
  and\ \bibinfo {author} {\bibfnamefont {S.}~\bibnamefont {Kurth}},\ }\href
  {\doibase 10.1063/5.0170312} {\bibfield  {journal} {\bibinfo  {journal} {J.
  Chem. Phys.}\ }\textbf {\bibinfo {volume} {159}},\ \bibinfo {pages} {154110}
  (\bibinfo {year} {2023})}\BibitemShut {NoStop}%
\bibitem [{\citenamefont {Liebert}\ \emph {et~al.}(2023)\citenamefont
  {Liebert}, \citenamefont {Chaou},\ and\ \citenamefont
  {Schilling}}]{Liebert2023_Refining}%
  \BibitemOpen
  \bibfield  {author} {\bibinfo {author} {\bibfnamefont {J.}~\bibnamefont
  {Liebert}}, \bibinfo {author} {\bibfnamefont {A.~Y.}\ \bibnamefont {Chaou}},
  \ and\ \bibinfo {author} {\bibfnamefont {C.}~\bibnamefont {Schilling}},\
  }\href {\doibase 10.1063/5.0143657} {\bibfield  {journal} {\bibinfo
  {journal} {J. Chem. Phys.}\ }\textbf {\bibinfo {volume} {158}},\ \bibinfo
  {pages} {214108} (\bibinfo {year} {2023})}\BibitemShut {NoStop}%
\bibitem [{\citenamefont {Senjean}\ \emph {et~al.}(2017)\citenamefont
  {Senjean}, \citenamefont {Tsuchiizu}, \citenamefont {Robert},\ and\
  \citenamefont {Fromager}}]{senjean2017local}%
  \BibitemOpen
  \bibfield  {author} {\bibinfo {author} {\bibfnamefont {B.}~\bibnamefont
  {Senjean}}, \bibinfo {author} {\bibfnamefont {M.}~\bibnamefont {Tsuchiizu}},
  \bibinfo {author} {\bibfnamefont {V.}~\bibnamefont {Robert}}, \ and\ \bibinfo
  {author} {\bibfnamefont {E.}~\bibnamefont {Fromager}},\ }\href
  {https://doi.org/10.1080/00268976.2016.1182224} {\bibfield  {journal}
  {\bibinfo  {journal} {Mol. Phys.}\ }\textbf {\bibinfo {volume} {115}},\
  \bibinfo {pages} {48} (\bibinfo {year} {2017})}\BibitemShut {NoStop}%
\bibitem [{\citenamefont {Lieb}(1983)}]{LFTransform-Lieb}%
  \BibitemOpen
  \bibfield  {author} {\bibinfo {author} {\bibfnamefont {E.~H.}\ \bibnamefont
  {Lieb}},\ }\href {https://doi.org/10.1002/qua.560240302} {\bibfield
  {journal} {\bibinfo  {journal} {Int. J. Quantum Chem.}\ }\textbf {\bibinfo
  {volume} {24}},\ \bibinfo {pages} {243} (\bibinfo {year} {1983})}\BibitemShut
  {NoStop}%
\end{thebibliography}

%

\end{document}